\pdfoutput=1

\newif\ifLogReview
\LogReviewtrue

\documentclass[
 aps,%
 pre,%
\ifLogReview
preprint,%
\else
reprint,%
\fi
 amsmath,
 amssymb,
 superscriptaddress,
longbibliography
]{revtex4-2}

\usepackage{
color,
xcolor,
graphicx,
hyperref
}

\hypersetup{pdfborder={0 0 0}, colorlinks=true,citecolor=blue,filecolor=blue,linkcolor=blue,urlcolor=blue}

\begin{document}


\title{
Clustering in atom probe tomography data: coordination number metric,
percolation-based parameter scaling and size effects
}

\author{Mykola Lazarev}
\affiliation{Akhiezer Institute for Theoretical Physics, 
National Science Center Kharkiv Institute of Physics and Technology, 
Akademichna 1, 61108 Kharkiv, Ukraine}
\email{n.lazarev@kipt.kharkov.ua}

\author{John Banhart}
\affiliation{Institute of Applied Materials, 
Helmholtz-Zentrum Berlin für Materialien und Energie GmbH, 
Hahn-Meitner-Platz 1, 14109 Berlin, Germany}
\email{banhart@helmholtz-berlin.de}


\begin{abstract}
The ability to identify nanometer-scale nuclei of new phases in atom 
probe tomography (APT) is often limited by the sensitivity of clustering 
algorithms to user-defined control parameters. Conventional approaches 
typically rely on the Euclidean distance metric and consider only solute 
atoms, thereby discarding the solvent atoms that contain most of the spatial 
information. Here, we introduce a coordination-number metric based on the
composition and apply it to higher-order clustering. Using various metrics, 
we investigate percolation in typical APT structures. By scaling 
clustering properties to the corresponding percolation thresholds, 
we define a self-similar variable that is almost invariant with 
respect to metrics, clustering parameters, and structural disorder. 
This variable provides a relevant description of clustering and enables 
the formal transfer of optimal parameters between clustering methods. 
We also study the characteristic clustering behavior in small precipitates 
and quantify how the precipitate-matrix interface alters the composition 
spectrum and broadens the clustering curve. 
Finally, using simulations that incorporate finite spatial resolution, 
detection efficiency, and other APT reconstruction artifacts, 
we show that the approach based on coordination numbers effectively 
compensates for heterogeneous dilations and outperforms
solute-density-based methods in all tested scenarios.
\end{abstract}


\maketitle

\section{Introduction}\label{Intro}
Understanding the mechanisms of precipitate formation and their
evolution is necessary for predicting the behavior of materials under
thermal and mechanical loading. Atom probe tomography (APT) is a unique
tool for acquiring such information at the atomic level \cite{Re1}. It
combines compositional sensitivity with position resolution, providing a
three-dimensional image of the chemical arrangement of up to
$10^9$ atoms. To reliably analyze the acquired dataset,
intrinsic limitations of APT must be considered, including detection
efficiency, spatial and chemical resolutions, and local magnification
effects \cite{Re2,Re3}.

Beyond materials characterization, APT data constitute a distinctive
class of disordered point patterns, making them attractive for studies
of percolation and cluster statistics. 
A note regarding terminology is necessary here. 
When describing the formation of new phases from a solid solution,
the following terms--often not clearly distinguished--are used: nuclei of
new phases, atom clusters, precipitates, etc. Atom clusters are typically
small and thermodynamically unstable, evolving into larger, more stable
precipitates. Conversely, in data analysis, atom positions are fixed,
and a set of parameters controls cluster identification. From this
perspective, a cluster is a network of atoms linked according to some
formal criteria. In this paper, connected groups of atoms identified by
cluster analysis will be called clusters, and the corresponding task
will be referred to as clustering; physical clusters will be referred to
as precipitates, and the process will be referred to as precipitation.

Characterization of large precipitates is usually straightforward. In 
this case, methods based on cells, concentration iso-surfaces, and
related methods are quite effective \cite{Re2,Re3}. 
However, small precipitates,
especially those with complex morphologies, require advanced clustering
techniques. Relevant approaches use correlations in the mutual
arrangement of solute atoms and include the maximum separation distance
(MSD) \cite{Re4}, density-based spatial clustering of applications with
noise (DBSCAN) \cite{Re5}, core-linkage \cite{Re6}, or Delaunay/Voronoi
tessellations \cite{Re7,Re8,Re9}. These methods use the spatial 
positioning of only solute atoms, which comprise a small fraction of the 
total number of detected atoms. Solvent atoms are typically involved 
in subsequent post-processing steps, such as enveloping and erosion. 
Hence, most APT data are disregarded in the initial, 
crucial step of clustering.

Defining the linkage between atoms is the cornerstone of cluster
analysis. Conventional methods, such as MSD, use the Euclidean distance
criterion. Recently, a maximum coordination number (MCN) method,
based on local composition, was proposed \cite{Re10}. 
Earlier, related approaches 
were also applied in APT data mining \cite{Re11,Re12,Re13,Re14}. 
A direct counterpart to MCN in machine learning is the $k$-nearest
neighbors (kNN) classifier, where $k$ plays the role of a coordination
number \cite{Re15}.

The coordination number of an atom is essentially the number of its nearest
neighbors. It is 12 in fcc and hcp crystals, and 8 in bcc crystals. In
glasses and liquids, it is defined by the first peak of the radial
distribution function \cite{Re16}. Effective coordination can be extended
to include next-nearest neighbors \cite{Re17}. For gas-type disorder, 
we can define local coordination using appropriate proximity criteria
\cite{Re18}.

Previously, we described the MCN method, constructed a map of phase 
compositions accessible for cluster analysis, and formulated a
criterion for selecting optimal control parameters by maximizing the
contrast between clustering degrees of different phases \cite{Re10}.
Those results focused on sufficiently large precipitates and neglected
interfacial contributions. Here, we extend MCN to higher-order
DBSCAN-type methods and focus on clustering in small precipitates, where
interfaces play the central role.

In what follows, we first describe the approach for simulating APT-like
structures, including the necessary parameters to account for
precipitate size and shape, limited detection efficiency, atomic spatial
resolution, and local magnification effects. We then formulate
coordination-number-based clustering methods that complement 
approaches based on the solute density. Next, we derive exact 
coordination-number correlation functions for the 
ideal solid solution and use them as the
basis for composition-spectrum analysis. We characterize clustering via
the percolation theory and show that the clustering behavior is governed by
a self-similar variable that combines composition with the methods'
control parameters. Using simulations of small, isolated precipitates,
we quantify the influence of the matrix--precipitate interface. 
Finally, we compare the performance of the clustering methods 
considered across various multiphase alloy states.

\section{Simulation setup}\label{Sect2}

The technique for simulating APT-like atomic structures has been detailed
in \cite{Re10}. The generation of random atomic coordinates starts from the
ideal fcc lattice with the lattice constant $L$. Periodic boundary conditions 
are used. Selected regions within the lattice are referred to as 
precipitates (P), approximated by spheroids with principal axes, 
$a = b$ and $c$. The remaining lattice forms the matrix (M). 
Both regions are assumed to be ideal solid solutions differing 
only in their solute concentrations $c_\text{M}$ and $c_\text{P}$. 
A solvent-only interfacial layer of thickness $L$ is located between 
the matrix and the precipitate to mimic 
the depletion zone surrounding precipitates.

Next, we remove part of atoms to simulate the limited detection
efficiency $\eta$. Uniform, uncorrelated atom removal is used in conventional
simulation practice \cite{Re19}. In APT, however, correlated losses
can occur near various inhomogeneities on the tip of the specimen \cite{Re20}. 
To address this effect, we remove a randomly selected atom along with its
neighbors at the $\chi - 1$ nearest sites, repeating this process until the 
target value of $\eta$ is reached. Obviously, at the correlated loss 
parameter $\chi = 1$ we simulate uncorrelated removal.

To simulate the given resolution $\Lambda$ of atomic coordinates we add
displacement vectors randomized in directions and magnitudes to the
coordinates so that the standard deviation is $\Lambda / 2$ along
each coordinate axis \cite{Re21}.

Finally, to simulate local magnification effects in APT \cite{Re3,Re22, Re23}
arising from phase heterogeneity and ambiguity of tip curvature, 
we deform the model crystal by applying a
triangular-wave transformation to the coordinates,
\begin{equation} \label{EQ_1_} 
\xi' = \xi + \delta\frac{\mathcal{L}_{\xi}}{2\pi}
\arcsin\left( \sin\frac{2\pi}{\mathcal{L}_{\xi}}\xi \right),\,\,\xi = x,y,z.
\end{equation} 
Here, $\mathcal{L}_{\xi}$ are the dimensions of the simulation box and
$\delta$ is the single strain parameter. This transformation results
in piecewise constant, alternating multiaxial strains, forming a
heterogeneous distribution of dilation and shear regions. 
All the parameters are listed in Appendix C.

The APT structures are typically characterized by the detection
efficiency $\eta  =  0.3 - 0.8$ and atomic coordinate resolution
$\Lambda  / L =  0.25 - 1$, where the lower limit of $\Lambda$
is reached only along certain directions \cite{Re3}. 
In this study, we fix $\eta$ to $0.5$. 
The parameter $\Lambda$ is set to $0.5, 1$, or $4$ in
units of the lattice constant $L$. The value $\Lambda = 0.5$ is
close to the best achievable one in APT, $\Lambda = 1$ corresponds to
the typical APT resolution for complex alloys, and $\Lambda = 4$
corresponds to the gas-type disorder. We mostly use $\chi$=1 and
$\delta = 0$, increasing $\delta$ up to $0.3$ to simulate density
heterogeneity.

We define the precipitate size by the average number of solutes
$n_\text{P}$ remaining after the random removal of atoms. Given the specified
precipitate solute concentration $c_\text{P}$, the total average number of
atoms in the initial precipitate is
$n_\text{P}  / \left( \eta c_\text{P} \right)$.

\section{Methods of cluster analysis}\label{Sect3}

Atomic links determine whether an atom in a solid solution belongs to a
particular cluster. Once the links have been established, cluster
identification is straightforward.

\subsection{Maximum separation distance}\label{Subsec31}

In the maximum separation distance (MSD) method \cite{Re4}, two solute
atoms are linked if their separation does not exceed the fixed cutoff
$d_{c}$. The connected network of linked solutes forms clusters.
Subsequently, solvent atoms within the $d_{c}$-neighborhood of each
solute may be assigned to the same cluster.

\subsection{Maximum coordination number}\label{Subsec32}

Let us consider the entire set of atoms, including both the solute and
the solvent. For each atom $i$ we form an ordered neighbor list of all
other atoms $j$, sorted by the proximity criterion e.g., by the Euclidean
distance. The coordination number $m_{ji}$ is defined as the sequence
number of the atom $j$ in the neighbor list of the atom $i$ and can
be regarded as the separation between atoms $j$ and $i$. However,
separations $m_{ji}$ and $m_{ij}$ often differ in disordered
structure, being unidirectional distances. That is, numbers $m_{ij}$
constitute a \emph{quasimetric} space \cite{Re24}. To obtain the true 
metric, we define the coordination distance as follows,
\begin{equation} \label{EQ_2_} 
M_{ij} = \frac{m_{ij} + m_{ji}}{2} .
\end{equation} 

Two atoms $i$ and $j$ are linked if $M_{ij} \leq z_{0}$, where
$z_{0}$ is a fixed cutoff. We refer to $z_{0}$ as the
maximum coordination number (MCN), 
like $d_{c}$ in the MSD method.
In an ideal crystal, $z_{0}$ can take values according to the number
of atoms in coordination shells (e.g., $z_{0} = 12,18,42$, etc. in
fcc), and each atom has exactly $z_{0}$ linked neighbors. However, in
a disordered system, at any $z_{0}$, the local number of neighbors
varies from atom to atom. The average number of neighbors $z$ depends on
the parameter $z_{0}$ and the degree of structural disorder. Fortunately,
the value of $z$ is only slightly less than $z_{0}$ for the metric 
\eqref{EQ_2_}.
Thus, the control parameter $z_{0}$ gives the average number of neighbors 
in MCN directly. The link structure generated by MCN is identical to
the \emph{k}-nearest-neighbor (kNN) graph \cite{Re18} when $k$ is set
equal to $z_{0}$. The distributions of local neighbor numbers for
different metrics are detailed in Appendix A.

Notice that unlike the Euclidean distance between two atoms, 
which relies solely on their individual coordinates, 
the coordination distance is inherently environment-dependent 
because all surrounding atoms determine it.

To establish the correspondence between the main parameters of MCN and
MSD, let us write the average neighbor number within the distance
$d_{c}$ as follows,
\begin{equation} \label{EQ_3_} 
z(d_{c}) = \rho_{0}\int_{0}^{d_{c}}{g_\text{rdf}(r)4}\pi r^{2} \, \mathrm{d}r,
\end{equation} 
where $g_\text{rdf}(r)$ is the radial distribution function,
$\rho_{0} = N / V$ is  average atomic density, $N$ is the total
number of atoms, and $V$ is the volume of the system.

\subsection{Coordination numbers based on the Voronoi diagram}\label{Subsect33}
\begin{figure}
\noindent
\ifLogReview
    \centering\includegraphics[width=0.5\textwidth]{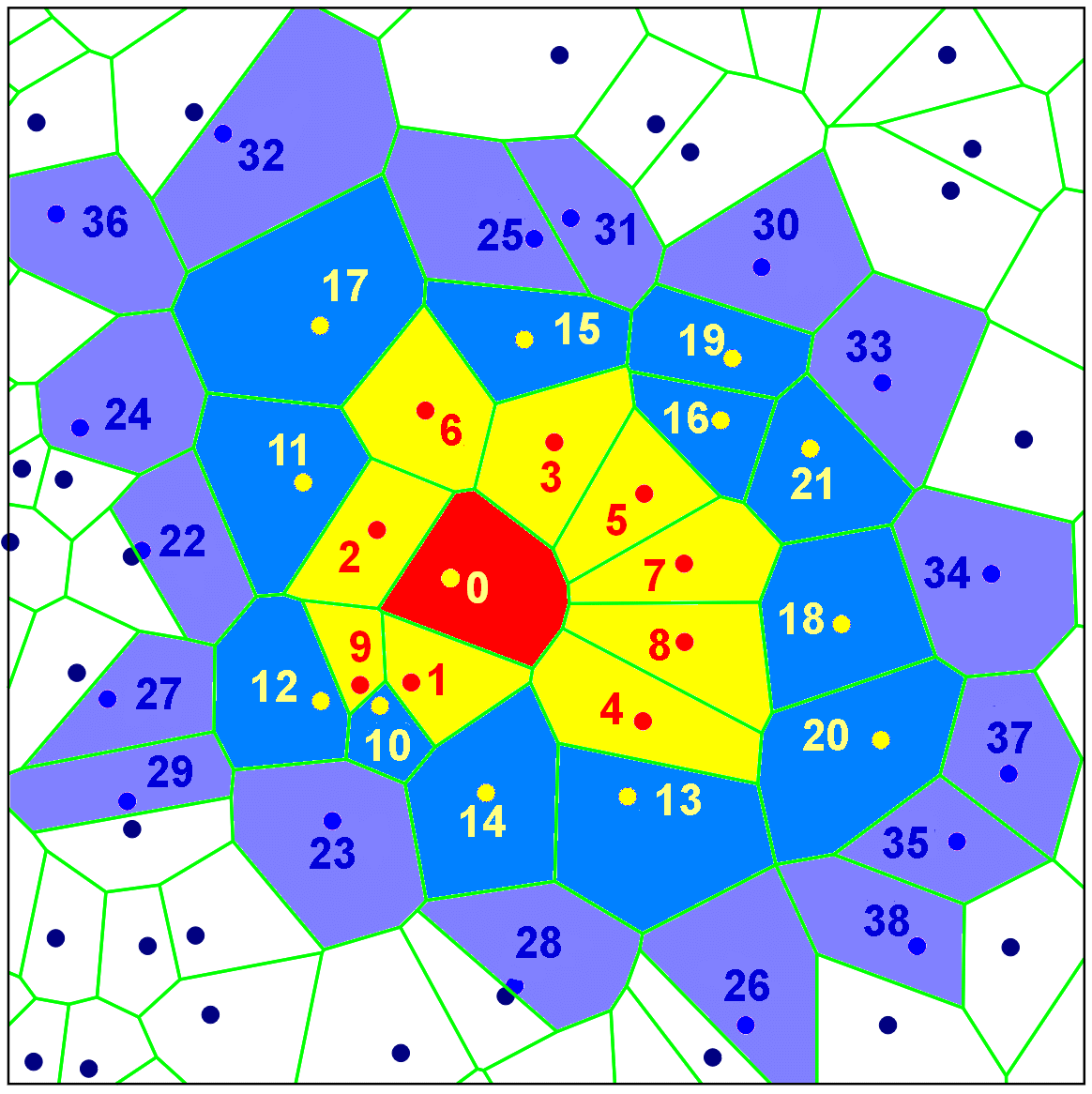}
\else
    \centering\includegraphics[width=0.3\textwidth]{image1.png}
\fi
\caption{
A two-dimensional example illustrating the definition
of coordination numbers based on the Voronoi diagram, shown as thin
lines. Neighbors in the first shell of atom $0$ are enumerated from $1$
to $9$ according to the values of the corresponding viewing angles of the
polygon edges. Neighbors in the second and third shells are numbered
according to the sum of such angles along the optimal path. }
\label{fig1}
\end{figure}

The atomic neighborhood can be determined not only by the Euclidean
distance criterion but also by the geometric relationship of atoms. Let
us show how coordination numbers can be derived from the Voronoi
diagram. Accordingly, two atoms are neighbors if they are separated by the
common face of touching Voronoi polyhedra. The neighbors of an atom
constitute the first shell. Each subsequent shell contains the outer
neighbors of the previous shell. The first five Voronoi's shells contain
$12, 42, 92, 162, 252$ atoms in the fcc lattice. In a disordered APT-like
structure, each atom has from four to several tens of nearest neighbors,
and the average numbers of atoms on the first five shells are about
$15.5, 70, 180, 360$ and $620$ \cite{Re25}.  
Figure~\ref{fig1} demonstrates the
definition of shells in a planar random structure, which was generated
with a relatively large spatial disorder parameter $\Lambda = 2$.
The Voronoi diagram was calculated using the software described 
in \cite{Re25b}. 

To enumerate the atoms in each separate shell, we take the solid angles 
related to the faces of polyhedral as the relevant proximity criterion.
This way, the neighbors in the first shell are numbered in descending
order of the corresponding solid angles. 
As an illustration in Fig.~\ref{fig1}, the atoms on the first shell are 
numbered according to the edges of the Voronoi polygon.

For each atom in subsequent shells, we find the optimal path that
maximizes the sum of the solid angles along the allowed paths, including
only links between atoms in neighboring shells. The method of coordination 
numbers based on the Voronoi diagram will be referred to as VCN. 
Representative maps of links obtained with different methods are given 
in Appendix A.

\subsection{Maximum coordination number of the $k$th order}\label{Subsect34}

The method of \emph{density-based spatial clustering of applications
with noise} (DBSCAN) groups densely packed atoms using the criterion of a
minimum of $k$ points in the epsilon neighborhood of the given point
\cite{Re5}. DBSCAN was first applied to APT data as a three-parameter
core-linkage clustering \cite{Re6}. Here, we formulate this method in the
coordination number representation.

Using higher-order correlations requires at least a two-step procedure.
First, we define the so-called core solute atoms. A solute atom
$p$ is a core one if within the coordination distance $z_{0}$ this
atom has no less than $k - 1$ neighboring solutes. Two parameters,
$k$ and $z_{0}$, define core solutes $p_\text{cor}$.

In the second step, we find the connectivity of the core atoms: two
core atoms, $p_\text{cor}$ and $q_\text{cor}$, are linked if their
coordination distance obeys
$M\left( p_\text{cor},q_\text{cor} \right) \leq z_\text{L}$, 
where $z_\text{L}$ is the
given linkage coordination distance. The usual clustering search is
performed to identify the cluster backbones.

Finally, all remaining solutes whose coordination distance does not
exceed $z_\text{L}$ from any core atom join the nearest core atom. The
core-linkage coordination-number method has three control parameters:
$k,z_{0}$, and $z_\text{L}$. In this paper, we restrict the method to the
two-parameter version, setting $z_\text{L} = z_{0}$ as in the initial
DBSCAN formulation \cite{Re5}. It is easy to verify that for $k = 1$ the
method reduces to the first-order MCN method, in which all solutes can
be treated as core atoms. We call the generalized method kMCN as it
originates from MCN. Accordingly, the $k$-order method based on the
Voronoi diagram is denoted as kVCN.

Notice that in the original DBSCAN, the $\varepsilon$-vicinity 
of a core point $p$ contains at least $k$ points, 
including point $p$, meaning the core point has no fewer 
than $k - 1$ neighbors. For compatibility, we
numerate the clustering order parameter $k$ accordingly. On the other
hand, the parameter $k$ is the sequence number of a neighboring solute
atom in the correlation function formalism. 
Thus, the correlation function of the $k$-th order corresponds to the 
kMCN clustering method of the $(k+1)$-th order.

\section{Correlation functions and composition spectrum analysis}\label{Sect4}
\subsection{Correlation functions}\label{Subsect41}
The pair correlation functions of the solute atoms complement the
corresponding cluster analysis method, see e.g. \cite{Re26}. A negative
binomial distribution describes the correlation function of the $k$th
solute atom on the coordination numbers $m$ in an ideal solid
solution of a given concentration $c$ \citep{Re10, Re27},
\begin{equation} \label{EQ_4_} 
B\left( m \middle| k,c \right) = 
\frac{(m - 1)!}{(k - 1)!(m - k)!}(1 - c)^{m - k}c^{k}, \,\, m \geq k.
\end{equation} 

To clarify the physical meaning of \eqref{EQ_4_}, we consider a related
correlation function in the coordinate representation, i.e. the
probability of finding the $k$th neighbor at the distance $r$ in a
gas-like structure \citep{Re6, Re28, Re29, Re30},
\begin{equation} \label{EQ_5_} 
P\left( r \middle| k,\alpha \right) = 
\frac{3\alpha r^{2}\left( \alpha r^{3} \right)^{k - 1}}{(k - 1)!}\exp\left( - \alpha r^{3} \right),
\end{equation} 
where $\alpha = 4\pi\rho /3$, $\rho = N_{B} / V$, and
$N_{B}$ is the number of solute atoms in the volume $V$.

A change of variables,
$m' = \left( 4\pi\rho_{0} / 3 \right)r^{3}$, reduces \eqref{EQ_5_} to the
gamma distribution \citep{Re27},
\begin{equation} \label{EQ_6_} 
G\left( m' \middle| k,c \right) = \frac{c \left( cm' \right)^{k - 1}}{(k - 1)!}\exp\left( - c m' \right),
\end{equation} 
where $\rho_{0} = N / V$ is the atomic density,
$c = N_{B} / N$ is the concentration of solutes, and
$N = N_{A} + N_{B}$ is the total number of atoms.

The relation~\eqref{EQ_6_} was obtained for a system with a gas-like coordinate
disorder. However, from a formal point of view, we can consider the
variable $m'$ in~\eqref{EQ_6_} as a quantity determined by \eqref{EQ_3_}, 
where $z$ and
$d_{c}$ are replaced by $m'$ and $r$ respectively. Since \eqref{EQ_3_} is
valid for an arbitrary homogeneous structure, the applicability of \eqref{EQ_6_}
extends far beyond the validity of~\eqref{EQ_5_}. 
The approximation~\eqref{EQ_6_} together
with~\eqref{EQ_3_} is much like the correlation function formalism used for an
interacting fluid \citep{Re31}. By this means the two-parametric ($k,c$)
discrete distribution~\eqref{EQ_4_} resembles 
the gamma distribution~\eqref{EQ_6_} with
respect to the continuous variable $m'$, and the correlation function
~\eqref{EQ_4_} converges to ~\eqref{EQ_6_} at $c \rightarrow 0$.

Consider the direct calculation of the correlation function in a
simulated or experimentally obtained system. Given the atom coordinates
we build ordered lists of all neighbors for each atom $i$. Let
$m_{ik}$ be the coordination number of the $k$th solute atom in
the neighbor list of the $i$th atom. Then the local
$k$th-order correlation function of atom $i$ is equal to 1 at
$m = m_{ik}$ and 0 otherwise, i.e.
$Q\left( i,k \middle| m \right) = \delta_{mm_{ik}}$, where
$\delta_{ij}$ is the Kronecker delta. The desired distribution
function is then found by averaging over the reference atoms $i$. 
Averaging can be done over either the solvent $i_{A}$, or solute
$i_{B}$ atoms. Both methods give the same result for a single-phase
solid solution. However, for a heterogeneous system, we prefer averaging
over solute atoms to achieve compatibility with coordinate-based
correlation functions and more reliable phase separation,
\begin{equation} \label{EQ_7_} 
Q_{B}\left( m \middle| k \right) = 
\frac{1}{N_{B}}\sum_{i_{B}}^{}{Q\left( i_{B},m \middle| k \right)}.
\end{equation}

\subsection{Composition spectrum analysis}\label{Subsect42}
A method for determining the local density spectrum of solute atoms was
previously proposed \citep{Re28,Re32}. The $k$-order correlation
functions \eqref{EQ_5_} were used as basis functions. A similar composition
spectrum analysis can be formulated in the coordination-number
representation. The idea is to expand the measured correlation function
\eqref{EQ_7_} into a series of basis functions \eqref{EQ_4_},
\begin{equation} \label{EQ_8_} 
Q\left( m \middle| k \right) = 
\sum_{j = 1}^{n}{g_{j}(k)B\left( m \middle| k,c_{j} \right)}.
\end{equation}
Here $c_{j}$ is the concentration and $g_{j}$ is the fraction of
solute atoms in the $j$th phase, $n$ is the number of phases. A
related density-based series were considered in \citep{Re6, Re30}.

Generalizing~\eqref{EQ_8_} is straightforward. Let $n$ tend to
infinity, and then change to the continuum limit,
\begin{equation} \label{EQ_9_} 
Q\left( m \middle| k \right) = 
\int_{0}^{1}{g(k,c)B\left( m \middle| k,c \right)\mathrm{d}c},
\end{equation}
where the function $g(k,c)$ can be viewed as the composition spectrum of
$k$th order. The spectral decomposition~\eqref{EQ_9_} is the 
Fredholm integral equation of the
first kind with respect to the function $g(k,c)$, and 
$B\left( m \middle| k,c \right)$ is its kernel. This equation is
ill-posed and requires special methods to solve it \citep{Re33, Re34}.

One might expect that for all $k$, the functions $g(k,c)$ should be
close to each other, but this is not always the case. The reason is that
these functions see the composition at different scales. The smaller the
$k$, the more detailed composition is detected. On the contrary,
increasing $k$ effectively suppresses small-scale variations in
composition.

\begin{figure}
\noindent
\ifLogReview
    \centering\includegraphics[width=0.75\textwidth]{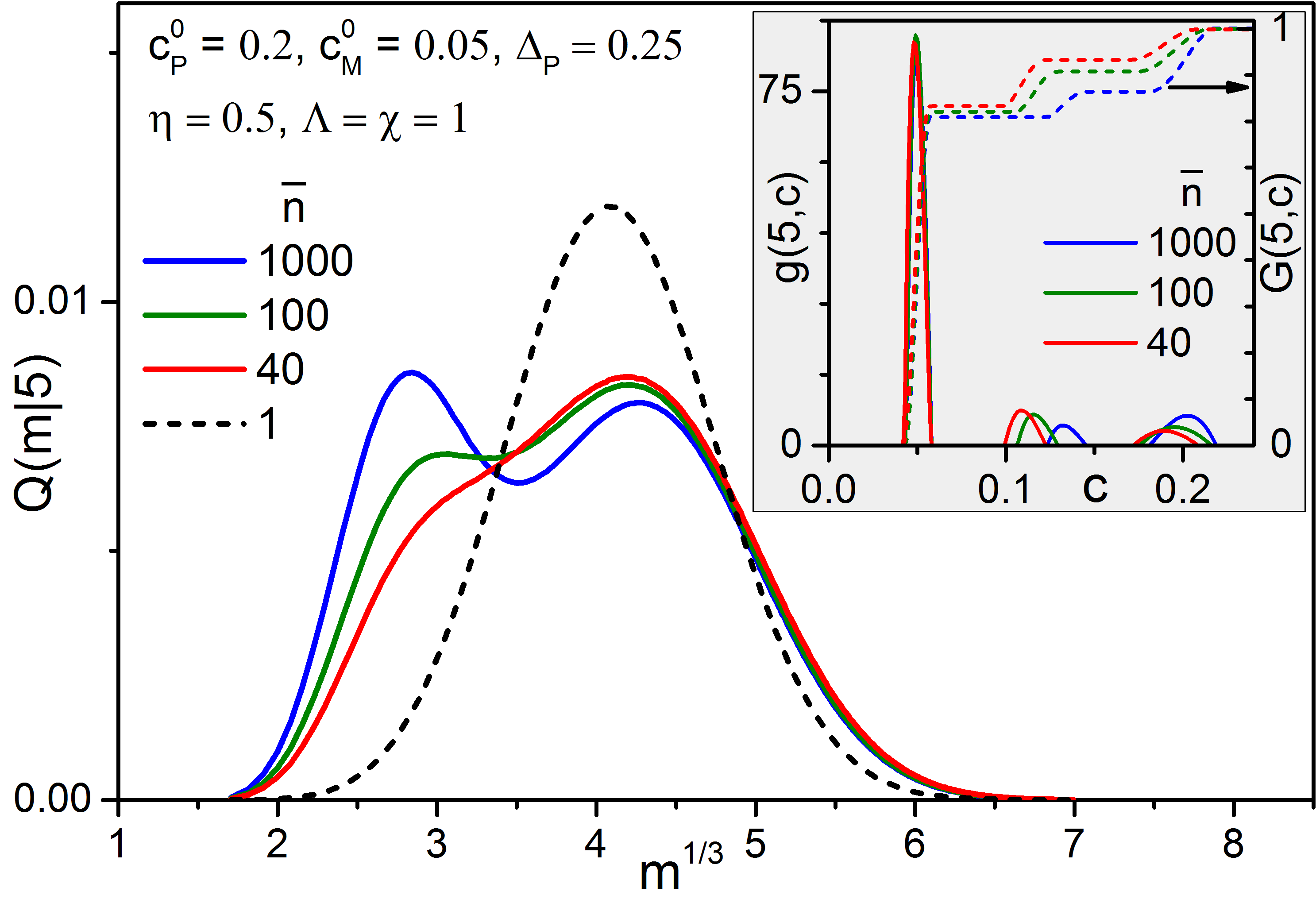}
\else
\centering\includegraphics[width=0.48\textwidth]{image2.png}
\fi
\caption{
Pair correlation functions
$Q\left( m \middle| k \right)$ of the 5th nearest solute in APT-like
structures with different sizes ($\overline{n}$) of precipitates are
shown by solid lines. The dashed line represents the corresponding
function found in a uniform solid solution of the same average
concentration of solutes. The inset shows the calculated composition
spectra $g(5,c)$ (solid lines) and the cumulative fractions of solutes
$G(5,c) = \int_{0}^{c}{g\left( 5,c' \right)dc'}$ (dashed lines).
}\label{fig2}
\end{figure}

To demonstrate the method, we simulate a set of two-phase
alloys containing precipitates of various sizes. The solute
concentrations in the matrix and precipitates are $c_\text{M} = 0.05$ and
$c_\text{P} = 0.2$, respectively; the fraction of solutes in the
precipitates is $\Delta_\text{P} = 0.2$; the lattice disorder parameters
are $\eta = 0.5, \Lambda = 1, \chi = 1$; and the sizes of the
precipitates are $\overline{n} = 40, 100$, or $1000$ solute atoms. The
calculated fifth-order correlation functions are shown as solid lines in
Fig.~\ref{fig2}. For the alloy with large precipitates ($\overline{n}=1000$)
the phase peaks are separated: the matrix peak is observed near the
coordination number $m = 78$, and the precipitate peak is around 
$m = 23$. As the precipitate sizes decrease, these peaks 
begin to converge and merge.

The solutions of the integral equation~\eqref{EQ_9_} 
for the considered correlation functions are shown by the solid lines 
in the insert to Fig.~\ref{fig2}.
The sharp, high peaks in the composition spectrum around
$c =  0.05$ correspond to the matrix and almost coincide at
different sizes $\overline{n}$. The right peaks near $c = 0.2$
correspond to the precipitate and the positions and heights of these
peaks depend on $\overline{n}$. Peaks in the intermediate region at
$c \in \lbrack 0.1, 0.15\rbrack$ correspond to the contribution
of the matrix-precipitate interfaces. For clarity, the cumulative
(integral) composition spectra are given as dashed lines in the insert
to Fig.~\ref{fig2} and directly show the contributions of various alloy
regions. For alloys with large precipitates
($\overline{n} = 1000$), the solute atom fractions in the matrix,
interfaces, and the bulk of the precipitates are $0.79, 0.06$, and
$0.15$, respectively. For alloys with small precipitates
($\overline{n} = 40$), these fractions are $0.81, 0.11$, and
$0.08$, respectively. In the latter case, more solute atoms are found
in the interface than in the bulk of the precipitates. As a
result, the effective solute concentration in small precipitates is
significantly lower than the nominal concentration. A more detailed
analysis shows that the interface contribution increases with decreasing
precipitate size $n$, increasing concentration $c_\text{P}$ and
coordinate disorder $\Lambda$.

\section{Percolation}\label{percolation}

Percolation theory describes clustering in random structures and is
widely used to understand various physical phenomena such as the
order-disorder phase transition in spin glasses, crack propagation in
structural materials, solidification of liquids, 
and other ones \cite{Re35, Re36}.
The properties of percolation in APT structures within the MCN method
were described in \cite{Re10}. Here, we provide a comparative study 
of the percolation observed in other methods.

\begin{figure}
\noindent
\ifLogReview
    \centering\includegraphics[width=0.75\textwidth]{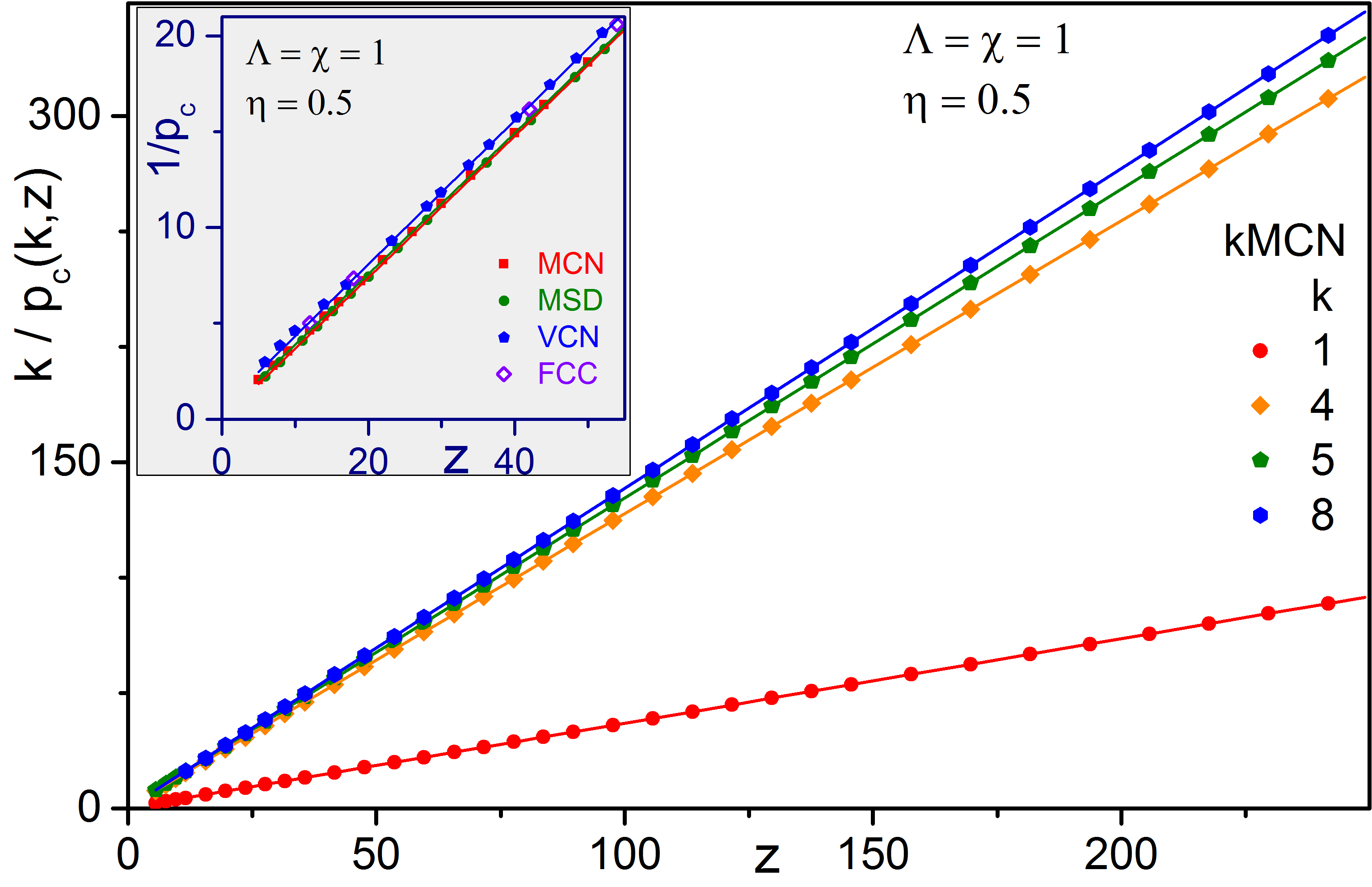}
\else
    \centering\includegraphics[width=0.48\textwidth]{image3.png} 
\fi
	\caption{
The correlations between the average coordination
number $z$ and the percolation threshold $p_{c}$ obtained by the
kMCN method at fixed $k$ in an APT-like structure. The lines show the
two-parameter fitting by~\eqref{EQ_11_}. 
The inset displays the same correlation obtained
using first-order clustering methods. Filled symbols show data in the
APT-like structure. Open symbols depict data in the fcc lattice with
extended-range neighborhoods~\cite{Re38}, where available coordination
numbers $z$ are $12, 18, 42$, etc. 
The lines show fitting by~\eqref{EQ_10_}.
}\label{fig3}
\end{figure}

The inset of Fig.~\ref{fig3} shows the dependencies of the percolation 
threshold $p_{c}$ on the average coordination number $z$ for 
different first-order clustering methods. 
The Newman-Ziff algorithm \cite{Re37} was
applied. The value of $z$ is calculated for the given $z_{0}$ and
specified structure. The relationship between $z$ and the MSD
clustering parameter $d_{c}$ is given by~\eqref{EQ_3_}. 
This figure also shows
the percolation thresholds on regular lattices \cite{Re38}, which are close
to those obtained by the VCN method in an APT-like structure.

For all considered methods, the value of $1 / p_{c}$ has a nearly
linear dependence on $z$ so that the following relation can
approximate the percolation threshold,
\begin{equation} \label{EQ_10_} 
p_{c}(z) = \frac{A}{z + a}.
\end{equation} 

Although the fitting parameters in~\eqref{EQ_10_} depend on the properties 
of disorder and clustering methods, the differences are small.

When discussing percolation in kMCN or DBSCAN, we use the term
"percolation in  $k$-order clustering models" here to avoid
confusion with the well-known $k$-core percolation ~\cite{Re36,Re39}.
The percolation thresholds in high-order models were obtained using a
modified algorithm since percolation occurs only via core atoms.
The modified algorithm is to some extent more complicated because 
at each step it requires verifying whether the added atom and its 
neighbors are core atoms. The modification increases processing
demand, yet it ensures the linearity of computation time with respect to
the total number of atoms. See more details in Appendix B.

Since the minimum cluster size is $k$ in kMCN, and percolation occurs
only through core atoms, the percolation threshold is expected to be
proportional to $k$. That is indeed the case, but only for large $k$. 
Figure~\ref{fig3} shows the correlation of $p_{c}$ and $z$ for kMCN.
We found that the percolation thresholds $p_{c}(k,z)$ at
$k = 2, 3$ almost coincide with those at $k = 1$. Then, starting
at $k = 4$, the values of $p_{c}(k,z)$ begin to increase with $k$.
For $k > 5$ the ratio $p_{c}(k,z) / k$ rapidly becomes weakly
dependent on $k$.

As Fig.~\ref{fig3} suggests, the inverse of the percolation threshold
$1 / p_{c}$ exhibits
almost perfect linearity with $z$ for all investigated values of
$k$. Thus, the direct generalization of~\eqref{EQ_10_} to the higher-order
methods reads
\begin{equation} \label{EQ_11_} 
p_{c}(k,z) = \frac{k A(k)}{z + a(k)}.
\end{equation}

The values of the fitting parameters $A(k)$ and $a(k)$ are presented
in Table~\ref{Tab1}. As mentioned above, $p_{c}(k) \approx p_{c}(1)$ at
$k = 2, 3$. Notably, the values of $A(k)$ obtained by different
clustering methods in different structures are almost identical to the
second digit after the decimal point. The values of the shift parameter
$a(k)$ differ slightly as well.
%
%
\begin{table}\centering
\caption{Fitting parameters of~\eqref{EQ_11_} and \eqref{EQ_12_} obtained by 
different methods in APT- and gas-like structures 
($\Lambda = 0.5-4, \eta = 0.5$), and the fcc lattice. The parameters were 
fitted in the range of $z_0$ from $6$ to $242$ for disordered structures 
and from $12$ to $248$ for the fcc lattice.}\label{Tab1}
\ifLogReview
\begin{minipage}{0.6\textwidth}
\else
\begin{minipage}{0.48\textwidth}
\fi
\begin{tabular*}{\columnwidth}{@{\extracolsep\fill}llcccccc@{\extracolsep\fill}}
\hline
Method  &  P$\backslash k$ & 1    & 4    & 5    & 6    & 7    & 8 \\
\hline
 kMCN          &  $A$     & 2.731 & 0.788 & 0.747 & 0.731 & 0.724 & 0.722\\
$\Lambda = 0.5$&  $a$     & 1.05  & 0.89  & 0.70  & 0.56  & 0.43  & 0.31\\
\hline
$\Lambda = 1$  &  $A$     & 2.729 & 0.787 & 0.747 & 0.731 & 0.724 & 0.721\\
               &  $a$     & 0.74  & 0.64  & 0.50  & 0.36  & 0.21  & 0.12\\
\hline
$\Lambda = 4$  &  $A$     & 2.734 & 0.789 & 0.748 & 0.731 & 0.724 & 0.721\\
               &  $a$     & 0.58  & 0.50  & 0.37  & 0.25  & 0.11  & 0.0\\
\hline
 DBSCAN        &  $A$     & 2.726 & 0.786 & 0.746 & 0.730 & 0.723 & 0.721\\
$\Lambda = 1$  &  $a$     & 0.25  & 0.21  & 0.18  & 0.15  & 0.13  & 0.11\\
 \hline
 fcc           &  $A$     & 2.730 & 0.787 & 0.746 & 0.730 & 0.723 & 0.721\\
               &  $a$     & 1.90  & 1.50  & 1.20  & 0.90  & 0.70  & 0.55\\ 
\hline
 kMCN          &  $A$     & 2.735 & 0.788 & 0.748 & 0.731 & 0.724 & 0.722\\
$\Lambda = 1$  &  $a$     & 1.22  & 0.90  & 0.77  & 0.56  & 0.40  & 0.35\\
Eq.~\eqref{EQ_12_} &  $B$ & -28.6 & -10.1 & -9.8 & -4.9 & -3.1 & -5.8\\
               &  $b$     & 40.2 & 19.0 & 17.1 & 6.2 & 0.4 & 4.7\\
\hline
\end{tabular*}
\end{minipage}

\end{table}

The correlation~\eqref{EQ_11_} is also observed in the ideal fcc lattice with
extended-range neighborhoods. In this case, the values of the factors
$A(k)$ almost coincide with those for the disordered structure, but
the parameters $a(k)$ appear to be $1.5$ to $0.5$ larger when
$k$ changes from $1$ to $8$, see Table~\ref{Tab1}. 
A larger value of $a(k)$ means
that the percolation thresholds in an ideal lattice are higher than in a
disordered structure at the given $z$. The same conclusion holds true when
comparing first-order methods.

To summarize, the slope parameter $A(k)$ can be roughly estimated as
$A(k) \approx e/k$, for $1 \leq k \leq 3$, and
$A(k) \approx e - 2$ for $k > 5$, where $e = 2.718$ is the Euler's
number. The asymptotes of $A(k)$ can be easily joined.

We can now make an estimate of the average number of neighboring solute
atoms at the percolation threshold, $N_{c}^{(s)}$. Since the solute
concentration is $N_{c}^{(s)}/z$, we immediately get from~\eqref{EQ_11_}
that for large $z$,
\begin{equation}\nonumber 
N_{c}^{(s)} \approx kA(k) .
\end{equation}

This estimation is nearly independent of $z$ and applies to both random
structures and regular lattices. It can help to select optimal
parameters for the DBSCAN method involving only solute atoms.

For a more precise description of the correlation between $p_{c}$ and
$z$, a second-order approximation can be used,
\begin{equation} \label{EQ_12_} 
p_{c}(k,z) = \frac{kA(k)}{z + a(k) + B(k) / \left( z + b(k) \right)}.
\end{equation}

The fitting parameters of~\eqref{EQ_12_} are given in Table~\ref{Tab1} as well. 
To compare the accuracies of~\eqref{EQ_11_}~and~\eqref{EQ_12_}, 
we consider kMCN with $k = 6$ for an
APT-like structure ($\Lambda = 1, \eta = 0.5$). In this case, the
relative error of the linear approximation~\eqref{EQ_11_} is about 3\% at
$z = 6$, decreasing to about 0.1\% at $z > 18$. The second-order
approximation~\eqref{EQ_12_} yields an error that is an order of magnitude
smaller. Its accuracy is even higher than that of the fifth-order
polynomial approximation of $1 / p_{c}(z)$ on $z$. Although the 
relation~\eqref{EQ_12_} has theoretical interest, 
the linear approximation~\eqref{EQ_11_} is sufficient for this work.

The relations~\eqref{EQ_11_} and \eqref{EQ_12_} indicate 
the existence of some self-similar variable composed of 
the average coordination distance $z$ and the solute
concentration $c$ at the fixed $k$. In a similar way to \citep{Re10} 
we introduce
\begin{equation} \label{EQ_13_} 
\zeta(k,z,c) = c / p_{c}(k,z),
\end{equation}
where $p_{c}(k,z)$ is defined by~\eqref{EQ_11_}. The normalization of $\zeta$ 
is chosen so that it is equal to one at the percolation threshold.

Let $\varphi(s,\zeta)$ be the size ($s$) distribution of solute
clusters in a random solid solution described by the 
average coordination number $z$ and solute concentration $c$ 
expressed in terms of the variable $\zeta$ defined by~\eqref{EQ_13_}. 
We assume that this distribution is normalized to
the total number of solute atoms as follows
$\sum_{s = 1}^{\infty}{s\varphi(s,\zeta)} = 1.$

The fraction of clustered solutes or the \emph{clustering curve} is
expressed by the complementary cumulative distribution function of
cluster size,
\begin{equation} \label{EQ_14_} 
F\left( s_{m},\zeta \right) = \sum_{s = s_{m}}^{\infty}{s\varphi(s,\zeta)}.
\end{equation}
It provides the probability of finding a solute atom in a cluster of at
least $s_{m}$ solutes.  This function helps select the main clustering parameters
and compare clustering methods~\citep{Re10}.
\begin{figure}
\noindent
\ifLogReview
    \centering\includegraphics[width=0.75\textwidth]{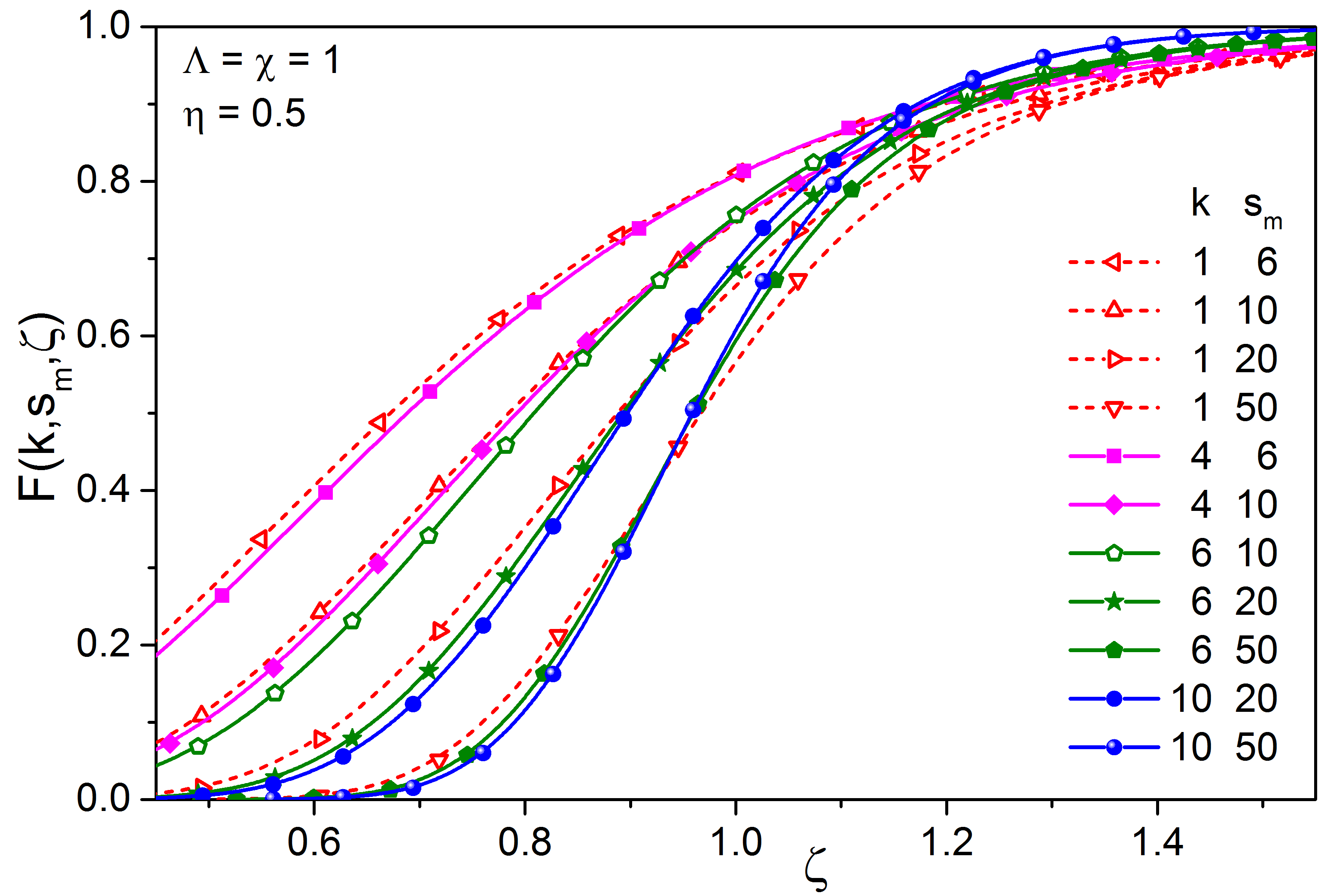}
\else
    \centering\includegraphics[width=0.48\textwidth]{image4.png}
\fi
\caption{
Clustering curves in the bulk depending on the control
parameters of the kMCN method.
} \label{fig4}
\end{figure}

Figure~\ref{fig4} demonstrates the behavior of $F\left( s_{m},\zeta \right)$
for kMCN at different control parameters $k$ and $s_{m}$. A comparison
between the first-order MCN and the fourth-order kMCN shows only minimal
differences, suggesting similar performance for these methods in cluster
analysis. As the order $k$ increases, the function
$F\left( s_{m},\zeta \right)$ pushes towards zero at $\zeta < 1$,
and it rapidly goes towards one at $\zeta >  1$. 
Due to such behavior, the high- and low-concentration phases 
can be effectively separated.

Examining the dependence of $F\left( s_{m},\zeta \right)$ 
on $\zeta$ (Fig.~\ref{fig4}) reveals
that it resembles a growth curve. The Gompertz function \cite{Re40}
appears to be the most appropriate one here \citep{Re10}.

\section{Simulation of finite size precipitates}\label{Sect6}
So far, we have explored clustering in a homogeneous infinite structure.
Real alloys contain several phases. Let the primary phase, called the
matrix, occupy most of the alloy with the concentration of solute atoms
$c_\text{M}$, and let the precipitates of the second phase be small and 
contain a higher concentration of solutes $c_\text{P}$. In such a model, 
all previously obtained results for the infinite structures are directly
applicable to the matrix. Precipitates require additional consideration.

\subsection{Clustering in isolated precipitates}\label{Subsect61}

The finite size of a precipitate changes the clustering behavior as the
influence of the matrix-precipitate interface increases. To estimate
this influence, we apply the formalism of the complementary cumulative
distribution function of cluster size \eqref{EQ_14_} for an ensemble of 
isolated precipitates of a given size, generated using the methodology 
described in Section~\ref{Sect2}, with $c_\text{M} = 0$. 
Then, we perform the cluster analysis using
the kMCN method with the fixed $k$, and vary the control parameter
$z_{0}$ over a wide range. The number of precipitates exceeds
$10^5$ in each run to ensure statistically reliable results.

\begin{figure}
\noindent
\ifLogReview
    \centering\includegraphics[width=0.75\textwidth]{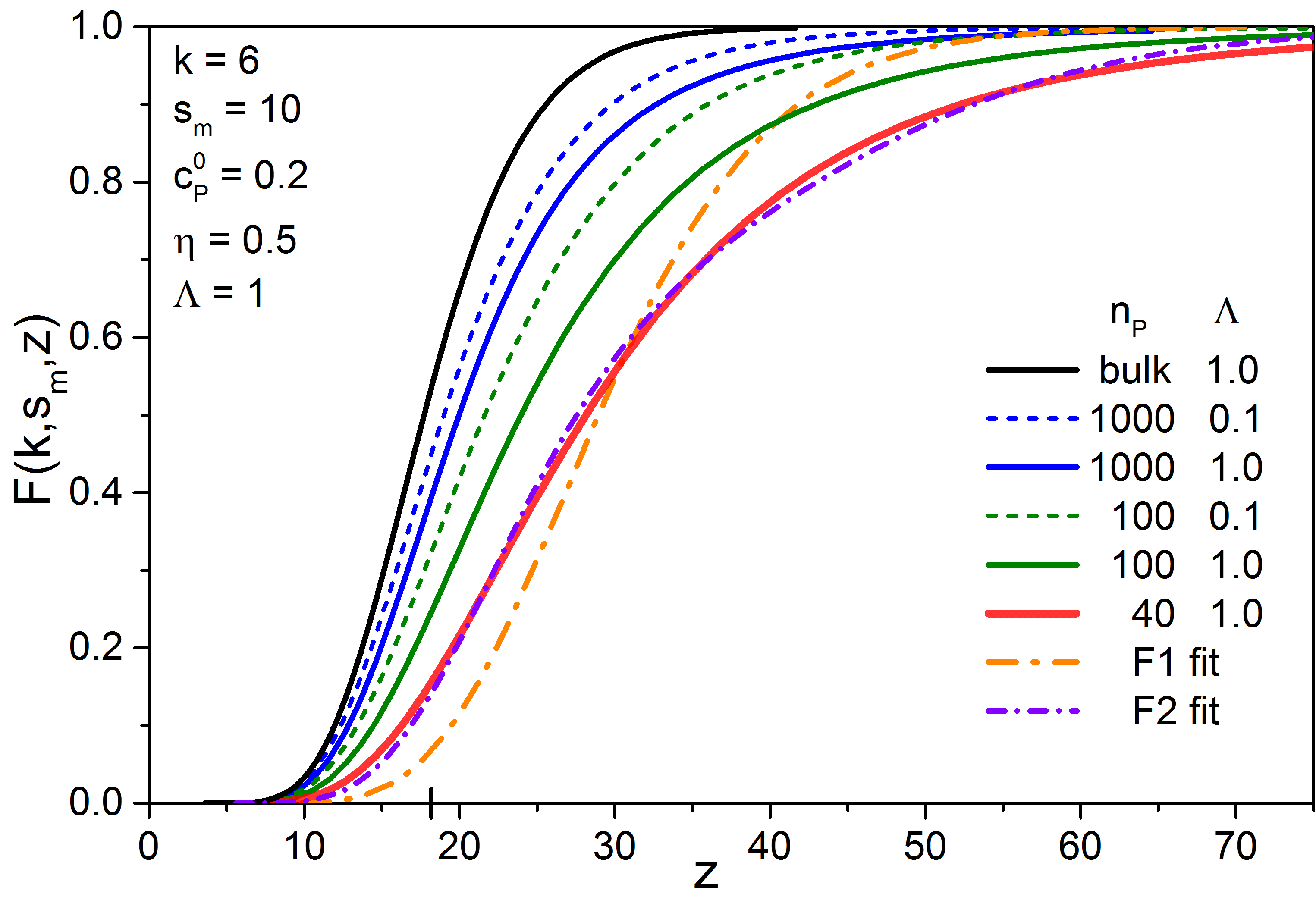}
\else
    \centering\includegraphics[width=0.48\textwidth]{image5.png}
\fi
\caption{Fraction of clustered solutes \emph{vs}. average
coordination number $z$ for isolated spherical precipitates of
different sizes $n_\text{P}$ and spatial resolution $\Lambda$. The
designed composition of precipitates $c_\text{P}^{0}$ and other parameters
are indicated in the plot. The dash-dot lines show the fitting of the
clustering curve for precipitates of size $n_\text{P} = 40$ by one- and
two-phase clustering models.} \label{fig5}
\end{figure}

Figure~\ref{fig5} shows the fraction of clustered atoms for different
precipitate sizes at the given nominal concentration of solute atoms
$c_\text{P}^{0} = 0.2$. The simulation conditions are indicated on the
graph. The leftmost curve corresponds to the infinite system. 
Since in this case the percolation occurs at about
$z = 18$ the clustered fraction becomes above $0.99$
for $z > 30$. In finite-size precipitates, the corresponding curves
are much flatter. For small precipitates ($n_\text{P} = 40$) almost
complete clustering ($F(z) > 0.99$) occurs only at $z > 90$, which
is three times more than in bulk clustering. This huge difference explains
why detecting small precipitates is only possible when the ratio
$c_\text{P} / c_\text{M}$ is high.

The clustering curves for $\Lambda = 0.1$ and $1$ are very close in
the bulk. However, in finite-size precipitates, these curves differ
noticeably, and the difference increases with decreasing precipitate
size, indicating a greater contribution of the interface layer as
$\Lambda$ increases.

The behavior of clustering curves in Fig.~\ref{fig5} suggests that the
apparent solute concentration is below the given value $c_\text{P}^{0}$. We
can select the effective concentration $c_{1}$, that makes the
clustering curve $F\left( k,s_{m},\zeta\left( z,c_{1} \right) \right)$
as close as possible to the observed one. 
The rightmost solid line shows the clustering curve for precipitates of size
$n_{P} = 40$. The long-dash-dot line shows fitting by a single-phase
system ($F_{1}$) with a concentration of $c_{1} = 0.115$. The
two-phase fitting ($F_{2}$) is significantly better, as shown by the
short-dash-dot line. The fitted concentrations are
 $c_{1} = 0.15$ and $c_{2} = 0.08$, and
the solute fraction in the second phase is $\Delta = 0.3$.

Figures~\ref{fig5}~and~\ref{fig2} confirm the validity of the 
two-phase precipitate model
in both composition-spectrum and clustering-curve analyses. However, the
fitting parameters $c_{1}$, $c_{2}$ and $\Delta$ are effective and
depend on the approach used.

\subsection{Optimality criterion for the parameters of the clustering method} \label{Subsect62}
Selecting optimal parameters requires an appropriate criterion. It can 
be directly defined for simulating data, as the properties of
precipitates are known in advance. Due to the limited detection
efficiency, the sizes of the simulated precipitates are
distributed in a wide interval, even if initially they were of the same
size. We compare the given size distribution of precipitates
$\varphi_{g}(s)$ with that resulting from cluster analysis
$\varphi(s,\lambda)$, where $\lambda$ is the set of parameters of a
clustering method, e.g., $\lambda \in k,z_{0},s_{m}\ $ for the kMCN. 
The closer these distributions are, the better the method's parameterization. 
Let us consider the following statistical distance
\begin{equation} \label{EQ_15_} 
\ifLogReview
W_{p}(\lambda) = \left(
\sum_{s = s_{m}}^{\infty}\left| s\varphi(s,\lambda) - s\varphi_{g}(s) \right|^{p} \middle/ 
\sum_{s = s_{m}}^{\infty}\left| s\varphi(s,\lambda) + s\varphi_{g}(s) \right|^{p}  
\right)^{1/p},
\else
W_{p}(\lambda) = \left\{ 
\frac{\sum_{s = s_{m}}^{\infty}\left| s\varphi(s,\lambda) - s\varphi_{g}(s) \right|^{p}}
{\sum_{s = s_{m}}^{\infty}\left| s\varphi(s,\lambda) + s\varphi_{g}(s) \right|^{p}} 
\right\}^{1/p},
\fi
\end{equation}
where the parameter $p$ takes the values $1$ or $2$.

The functional \eqref{EQ_15_} can be viewed as a criterion for selecting good
clustering control parameters. It is positively definite and symmetric
with respect to the distributions $\varphi(s,\lambda)$ and
$\varphi_{g}(s)$. It vanishes when the distributions coincide and
reaches $1$ when strongly uncorrelated. The minimum of $W_{p}(\lambda)$
determines the optimal parameter set $\lambda$. The criterion \eqref{EQ_15_} 
for both values of $p$ gives the same optimal parameter set $\lambda$
only if the given and calculated distributions are close to each other,
implying that $W_{p}(\lambda) \ll 1$.
Notice that criteria \eqref{EQ_15_} are only used to determine the optimal
set of parameters $\lambda$ and to compare different clustering
methods for a given $\varphi_{g}(s)$.

\subsection{Clustering in a two-phase system}\label{Subs63}
\subsubsection*{First-order methods}
\begin{figure*}
\noindent
\ifLogReview
\centering (a)\\
    \centering\includegraphics[width=0.75\textwidth]{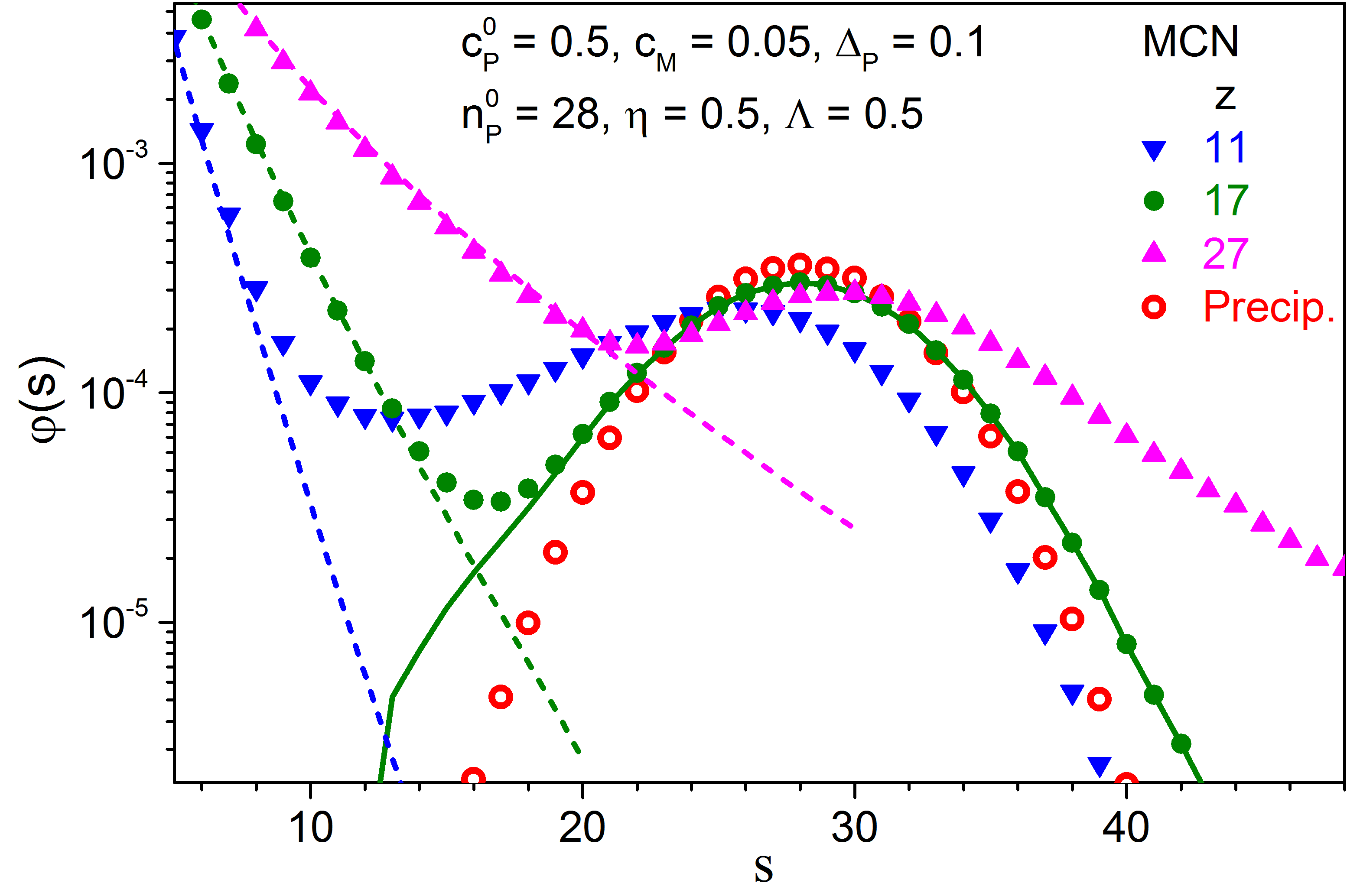}\\
(b)\\   
	\centering\includegraphics[width=0.75\textwidth]{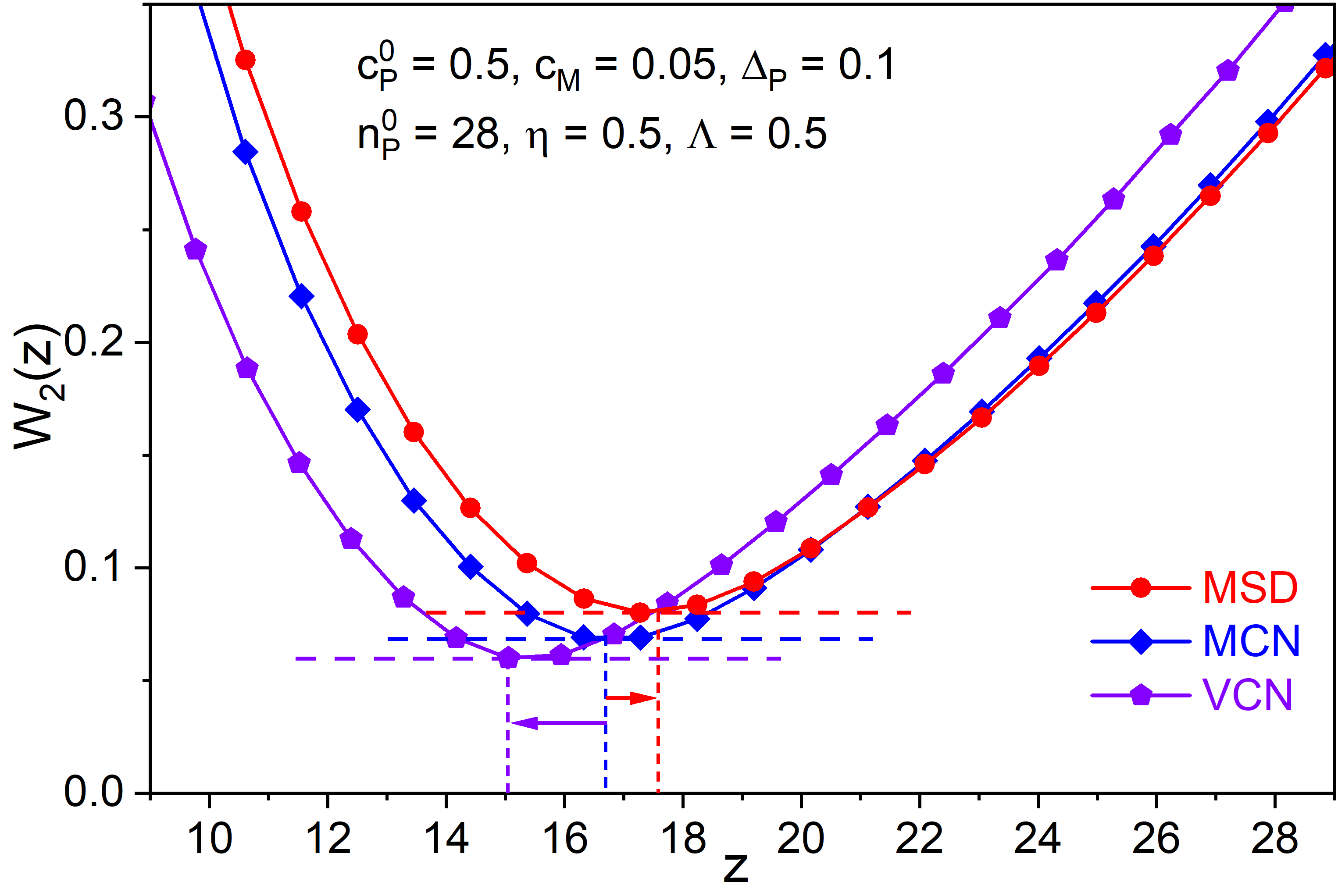}
\else
\centering (a)\hspace{25em} (b)\\
    \raggedleft\includegraphics[width=0.48\textwidth]{image6.png} 
\hfill    
	\raggedright\includegraphics[width=0.48\textwidth]{image7.png}
\fi
\caption{
(a) Cluster size distributions obtained by the MCN method for
three values of the parameter $z$ at the concentration ratio
$c_\text{P}^{0} / c_\text{M} = 10$. Open circles display the given size
distribution of precipitates. 
Filled symbols show calculated cluster distributions; 
dashed lines show the corresponding distributions in the matrix; 
the green solid line shows the optimal reduced distribution 
defined by \eqref{EQ_16_}. 
(b) Comparison of the efficiency of cluster analysis using different
first-order clustering methods. The dashed lines show the positions of the
$W_{2}(z)$ minima, which define the optimal $z^{\text{(opt)}}$ of the
methods.
} \label{fig6}
\end{figure*}

Figure~\ref{fig6}a shows the cluster-size distributions calculated by the MCN
method for the simulated two-phase system with the matrix concentration
$c_\text{M}$ and relatively high precipitate concentration $c_\text{P}$.
About $30$ structure realizations were generated, yielding more than $10^5$ 
precipitates. Then, a cluster analysis was performed for the control
parameter $z$, ranging from $8$ to $48$ in steps of $1$. Three of the
resulting distributions are shown.

Both criteria \eqref{EQ_15_} yield the optimal parameter $z^{\text{(opt)}}$ 
of about $17$ (filled circles). Two distributions obtained using non-optimal
parameters are also shown for comparison. At optimal parametrization,
three characteristic intervals of cluster sizes can be distinguished.
Background clusters from the matrix
$\widehat{\varphi}\left( s,z,c_\text{M} \right)$ dominate at small sizes
($s < 15$), as shown by dashed lines in Fig.~\ref{fig6}. The matrix
contribution can be subtracted,
\begin{equation} \label{EQ_16_} 
\varphi_{R}(s,z) = \varphi(s,z) - 
(1 - \Delta_{P})\widehat{\varphi}\left( s,z,c_\text{M} \right),
\end{equation}
where $\Delta_{P}$ is the fraction of solutes in precipitates. The
reduced distributions are shown by the solid line. In the
intermediate range ($15 < s < 35$), a good agreement with the precipitate
 distribution is observed. The tail of the distribution for $s > 35$
is due to the attachment of matrix solutes to clusters in precipitates.
This effect is particularly evident when the control parameter is
overestimated, as for $z = 27$ in Fig.~\ref{fig6}a. Conversely, an
underestimated $z = 11$ does not ensure complete clustering of the
precipitate region.

The cluster distributions in Fig.~\ref{fig6}a, obtained using the MCN method,
appear similar to those from other clustering methods. To make a
quantitative comparison, we use the $W_{2}(z)$ metric for various
methods, with $d_{c}$ and $z$ related by \eqref{EQ_3_}. 
As shown in Fig.~\ref{fig6}b,
the optimal values of the control parameter $z$ differ across the
methods under consideration, and the $W_{2}(z)$ curves are shifted
slightly to the right and left for the MSD and VCN methods,
respectively. This behavior reflects the dependencies of the percolation
thresholds, shown in Fig.~\ref{fig3}.
\begin{figure}
\noindent
\ifLogReview
    \centering\includegraphics[width=0.75\textwidth]{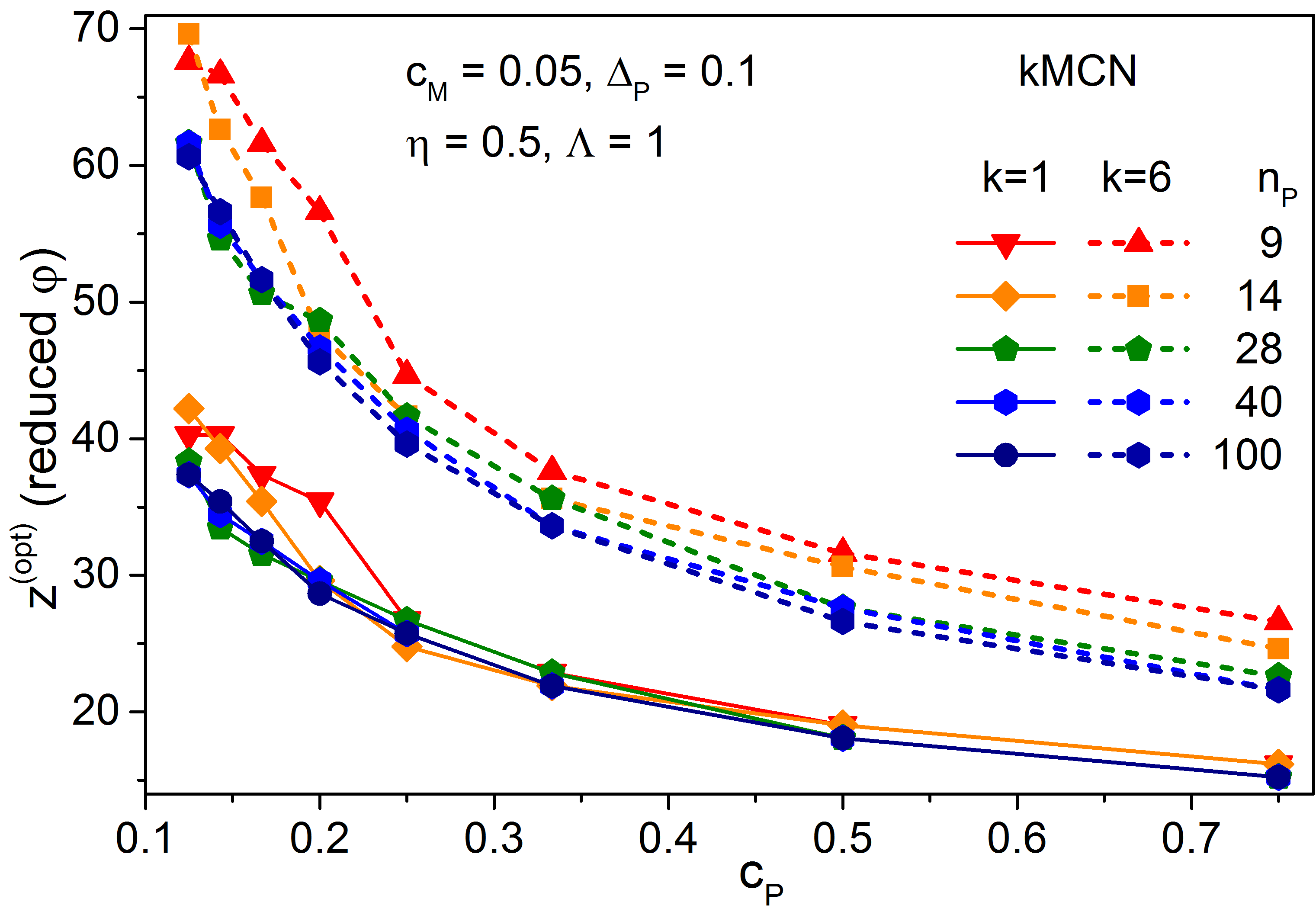}
\else
    \centering\includegraphics[width=0.48\textwidth]{image8.png}
\fi
\caption{
Dependencies of the optimal control parameter 
on the composition and size of the precipitates in kMCN methods after
reduction of the matrix contribution according to \eqref{EQ_16_}.
} \label{fig7}
\end{figure}

The optimal value of the coordination number $z^{\text{(opt)}}$ depends on
the size and composition of precipitates, as seen in Fig.~\ref{fig7}. 
For small precipitates, the optimal $z^{\text{(opt)}}$ is slightly increased,
due to the contribution of the precipitate-matrix interface, 
see Figs.~\ref{fig2}~and~\ref{fig5}.
However, the dependence on the precipitate size is small.

The compositional dependence $z^{\text{(opt)}}(c_\text{P})$ appears to be
$\propto 1 / \sqrt{c_\text{P}}$ , as shown in Fig.~\ref{fig7}. This behavior
can be understood based on the requirement to select the $z^{\text{(opt)}}$ 
value that ensures a high degree of clustering in the precipitate and minimal
clustering in the matrix. This degree of clustering is controlled by the
self-similar variable, $\zeta \propto z c$, which must be much greater
than one in the precipitate and much less than one in the matrix. 
Given the sharp dependence of $F(\zeta)$ in Fig.~\ref{fig5}, 
we obtain the heuristic relationship 
$z^{\text{(opt)}} \propto {{(c}_\text{P}c_\text{M})}^{- 1/2}$.

\subsubsection*{High-order methods}
\begin{figure}
\noindent
\ifLogReview
    \centering\includegraphics[width=0.75\textwidth]{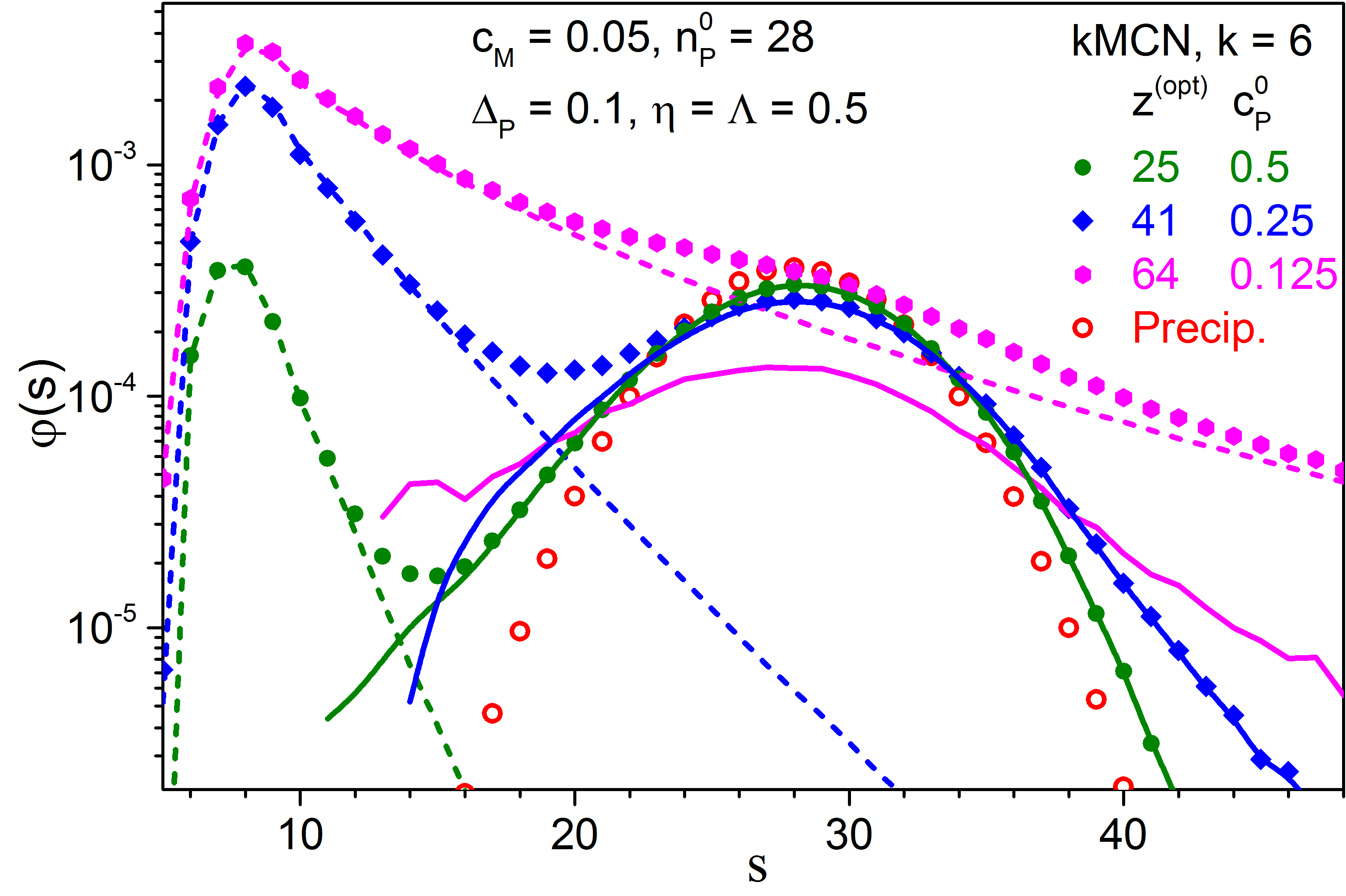}
\else
    \centering\includegraphics[width=0.48\textwidth]{image9.png} 
\fi
\caption{
Cluster size distributions obtained by the kMCN
method at $k = 6$ for different precipitate compositions at optimal
control parameters. The random structure property and labelling are the
same as in Fig.~\ref{fig6}a. 
} \label{fig8}
\end{figure}

Figure~\ref{fig8} presents the optimal cluster size distributions for the kMCD
method at $k = 6$. At the precipitate composition
$c_\text{P} = 0.5$, the system is identical to that shown in 
Fig.~\ref{fig6}, where the selection of the optimal 
control parameter of first-order methods was demonstrated. 
We see two main differences comparing dependencies in 
Figs.~\ref{fig8}~and~\ref{fig6}. 
The first one is the increase in the optimal control parameter
 $z^{\text{(opt)}}$ by about $1.5$. The second is
the significant suppression of the low-sized tail in the unreduced
distributions. However, the reduced distributions, shown as solid lines
in both figures, bear a striking resemblance.

Figure~\ref{fig8} also demonstrates how the quality of the optimal cluster size
distributions changes with the $c_\text{P}^{0} / c_\text{M}$ ratio. At the 
ratio of 10 ($c_\text{P}^{0} = 0.5$), the agreement with the precipitate
size distribution is mostly very good. At $c_\text{P}^{0} = 0.25$, the
deviation from the specified precipitate distribution is notable.
However, by carefully selecting the control parameters, the
reduced distribution yields the correct fraction of clustered solute
atoms. Finally, at $c_\text{P}^{0} = 0.125$, the resulting cluster
distribution is far from the specified precipitate distribution, although 
a deviation from the randomized arrangement of solute atoms  can be detected.

A quantitative comparison shows a slight advantage of high-order methods
at the most problematic small $c_\text{P} / c_\text{M}$ ratios.
Surprisingly, in the opposite case of a large $c_\text{P} / c_\text{M}$
ratio, the first-order methods have a
slight advantage according to the $W_{2}(\lambda)$ criterion at
optimal values of $z$ or $d_{c}$. This behavior is related to the
features of high-order methods resulting in smoothing the local
composition inhomogeneities. The same features are observed in the 
correlation functions (Fig.~\ref{fig8}). 
The interface boundary appears to be effectively broadening.
This effect is more pronounced for small precipitates, limiting the applicability 
of $k$-order methods to precipitate sizes roughly $n_\text{P} > 2k$.

A comparison of the optimal parameters' dependencies, 
shown in  Fig.~\ref{fig7}, is also noteworthy. 
First, we observe a similarity between the
results obtained by methods of different $k$. However, in addition to
the obvious scaling to large values of the control parameter $z$,
high-order $k$ methods show a more pronounced dependence on the
precipitate size $n_\text{P}$ due to the influence of the
precipitate-matrix interface.
\begin{figure}
\ifLogReview
    \centering\includegraphics[width=0.75\textwidth]{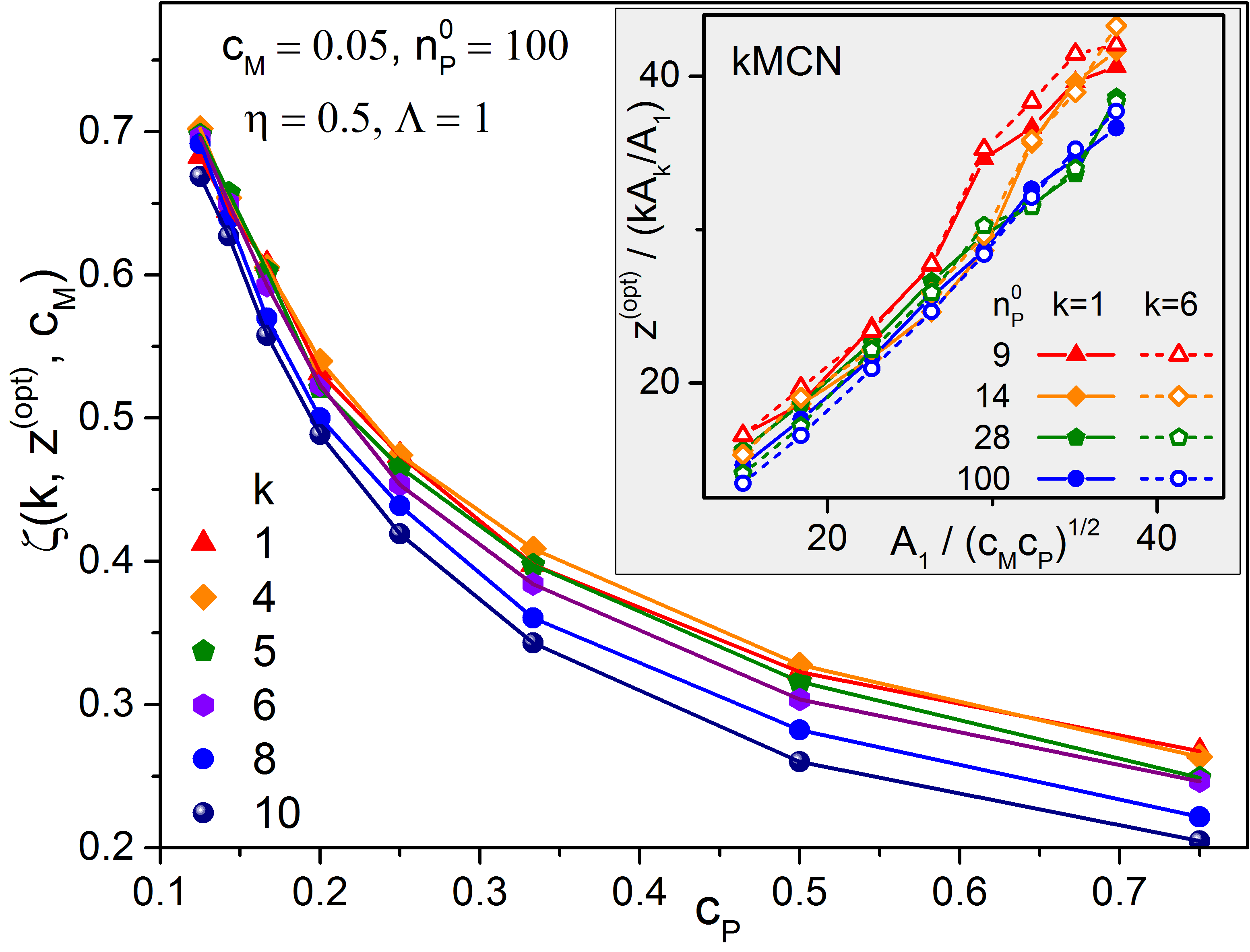}
\else
    \centering\includegraphics[width=0.48\textwidth]{image10.png}
\fi
\caption{
The dependence of the optimal self-similar variable $\zeta^{\text{(opt)}}$
in the matrix for different order parameters $k$ of kMCN methods. 
The inset figure shows the scaling of the optimal parameter $z^{\text{(opt)}}$ 
with the method order $k$ and the precipitate composition $c_\text{P}$.
} \label{fig9}
\end{figure}

To describe the scaling of $z^{\text{(opt)}}$ values with the method order
$k$ quantitatively, we present the dependence of the optimal self-similar
variable $\zeta^{\text{(opt)}} \equiv \zeta(k,z^{\text{(opt)}},c_\text{M})$ 
defined by
\eqref{EQ_13_} on the precipitate composition, as shown in Fig.~\ref{fig9}.
We see almost perfect scaling for $1 \leq k \leq 10$. With increasing
$k$ and the ratio $c_\text{P} / c_\text{M}$, there is some downward
deviation of $\zeta^{\text{(opt)}}$ values. However, since the optimality
region for $\zeta^{\text{(opt)}}$ widens with the increase in the
$c_\text{P} / c_\text{M}$ ratio, this deviation remains insignificant.

Combining the data shown in Figs.~\ref{fig7}-\ref{fig9}, we are convinced that
$z^{\text{(opt)}}$ is inversely proportional to 
${{(c}_\text{P}c_\text{M})}^{1/2}$
for arbitrary $k$, at least for the large precipitates; 
see the inset in Fig.~\ref{fig9}.

\begin{figure}
\noindent
\ifLogReview
    \centering\includegraphics[width=0.75\textwidth]{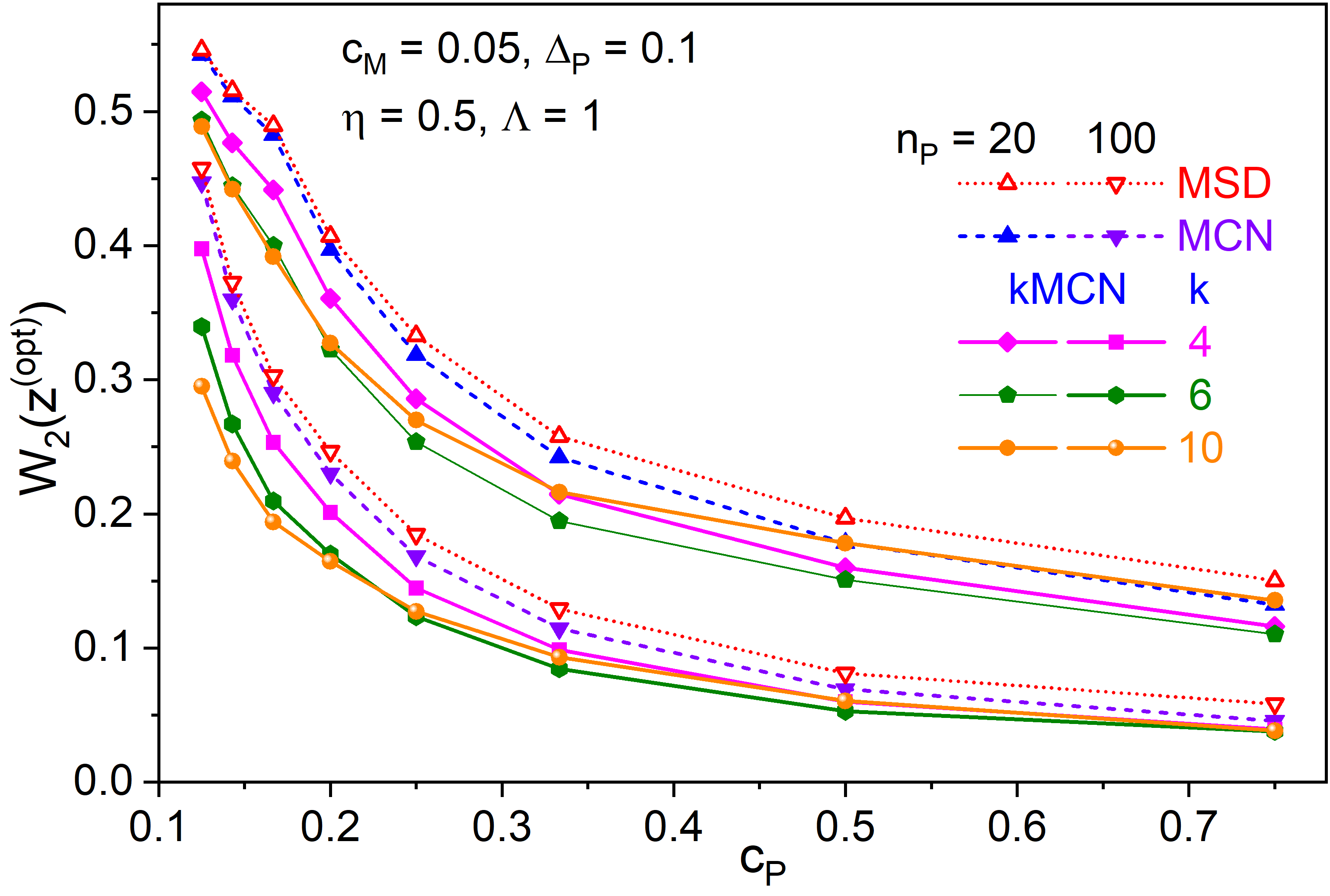}
\else
    \centering\includegraphics[width=0.48\textwidth]{image11.png}
\fi
\caption{
Comparison of quality metrics \eqref{EQ_15_} obtained by different
methods depending on the composition and size of spherical precipitates.
} \label{fig10}
\end{figure}

Figure~\ref{fig10} provides a quantitative comparison of clustering quality 
as a function of the method order $k$, as well as the size and composition of 
precipitates. The MCN method consistently outperforms the MSD method. 
Although not shown in the figure, DBSCAN is also less efficient than the 
corresponding kMCN method.
The results for small and large precipitates are clearly separated, and the 
clustering quality for large precipitates is higher (corresponding to a lower
$W_{2}$ value). The dependence of $W_{2}$ on the method order $k$ is
more complex. At a low $c_\text{P} / c_\text{M}$ ratio, increasing the value 
of $k$ positively affects the clustering quality. However, at high
$c_\text{P} /c_\text{M}$, a non-monotonic dependence of $W_{2}(k)$ 
is observed, with the minimum near $k = 6$. These observations are valid only
for $n_\text{P} \gtrsim 2k$.

\subsection{Impact of local magnification effects on the
performance of clustering methods}\label{Subsect64}

Due to the heterogeneity of the sample composition, distortions of the
trajectories of evaporated atoms occur, manifested in local
magnification effects \cite{Re22}. Ambiguity in the tip curvature, crystal
defects, and poles can also lead to observed density heterogeneities. An
approach to simulating magnification effects by local anisotropic
deformation of the precipitate region was described in \citep{Re23}. We
employ a similar, yet simpler, technique, applying non-uniform
deformation according to \eqref{EQ_1_}.

\begin{figure}
\noindent
\ifLogReview
    \centering\includegraphics[width=0.75\textwidth]{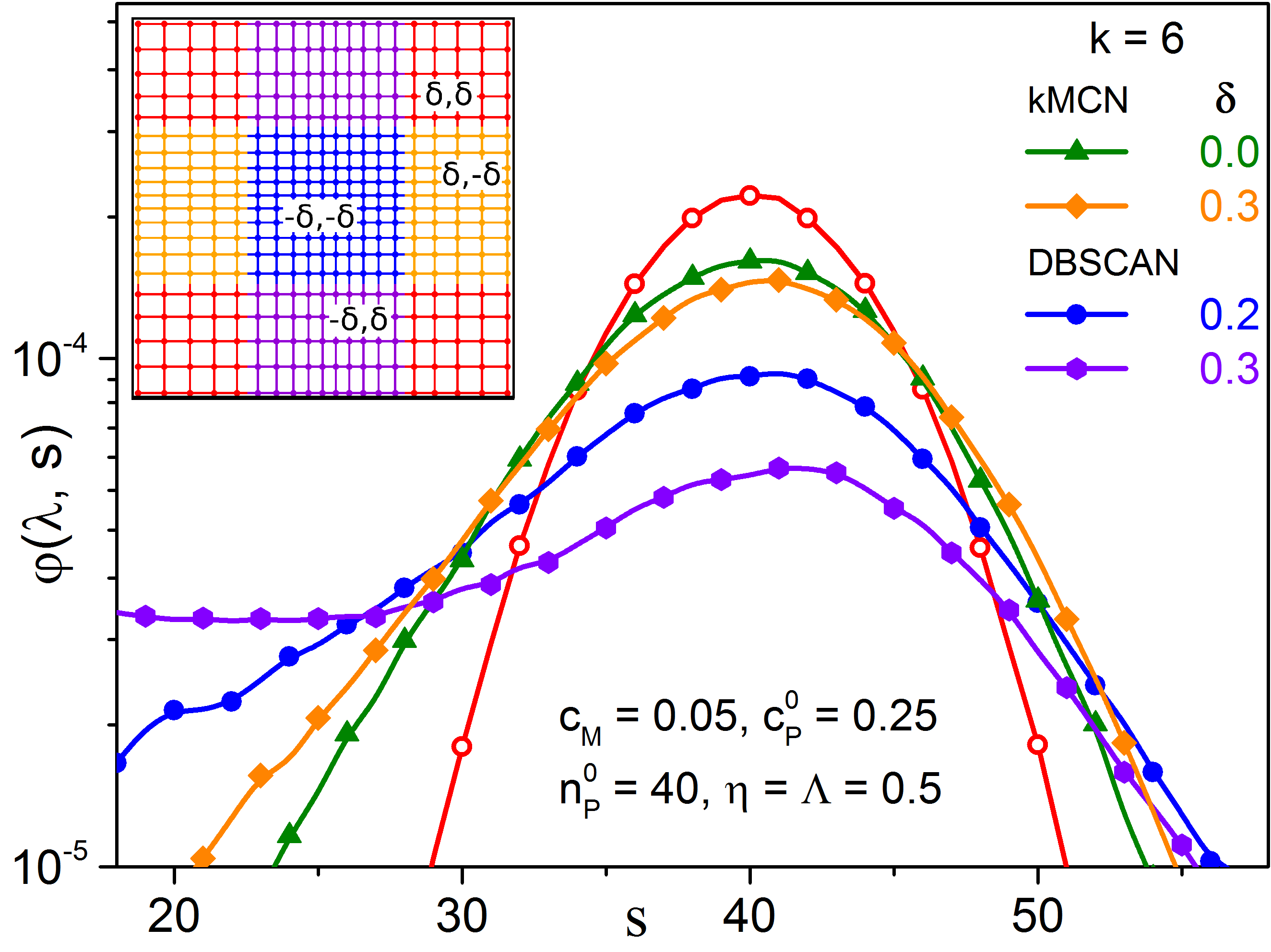}
\else
    \centering\includegraphics[width=0.48\textwidth]{image12.png}
\fi
\caption{
The impact of heterogeneous strain on the results of
cluster analysis obtained by kMCN and DBSCAN. Cluster size distributions
at optimal control parameters are shown at several values of the strain
parameter $\delta$. The random structure properties are indicated on
the plot.
The insert: Two-dimensional illustration of piecewise constant
deformation \eqref{EQ_1_} of square lattice at the strain parameter 
$\delta = 0.3$. Circles and lines indicate atom positions 
and interatomic links, respectively. 
} \label{fig11}
\end{figure}

The features of the applied deformation can be illustrated by a
two-dimensional lattice, as shown in the insert to Fig.~\ref{fig11}, 
where four types of deformed regions are distinguished,
corresponding to diagonal elements of the strain tensor,
${\widehat{\varepsilon} \equiv (\varepsilon}_{xx},\varepsilon_{yy}) 
= ( \pm \delta, \pm \delta)$.
Regions of uniform expansion and compression are observed, 
as well as regions of shear. Since the two shear regions are equivalent, 
three types of deformation are represented in this lattice.

The strain tensor in a 3D lattice, as defined in \eqref{EQ_1_}, also has 
only nonzero diagonal elements and provides four types of deformed areas. 
One-eighth of the crystal is uniformly compressed, and the same part is
uniformly expanded. The remaining parts are mixed strained, e.g.,
${\widehat{\varepsilon}}^{*} = \pm (\delta,\delta, - \delta)$, and contain 
both shear and dilatation components. The shapes of precipitates in regions 
with mixed strain change from spherical to ellipsoidal.

Figure~\ref{fig11} compares the sensitivity of the methods to non-uniform 
deformation. Even the smallest strain $\delta$ negatively affects the 
performance of DBSCAN, whereas the kMCN method 
demonstrates greater robustness to this type of deformation. As $\delta$ 
increase, some degradation in the quality of the cluster analysis using 
the kMCN method is attributable to the shear component of the strain.

Another advantage of the kMCN method is that the optimal value of the 
control parameter $z^{\text{(opt)}}$, equal to $40$ in this case, does not
depend on the strain parameter $\delta$.
On the other hand, the optimal DBSCAN parameter $d_{c}^{\text{(opt)}}$  
decreases from $0.68$ to $0.58$~\emph{nm} as $\delta$ 
increases from $0$ to $0.3$.

The kVCN method, based on Voronoi diagrams, demonstrates even greater 
resistance against strain-induced distortions. However, these benefits are 
evident only at low levels of noise in the atomic coordinates $\Lambda$. 
As the noise level increases, the advantages of kVCN vanish, and parity is 
reached between kVCN and kMCN methods.

\subsubsection*{Effects of pure dilatation and pure shear on clustering}

Given the results presented in Fig.~\ref{fig11}, the origin of this 
behavior remains unclear. When applying the strain field 
\eqref{EQ_1_}, some areas undergo pure uniform compression or 
expansion with local volume changes by factors of 
${(1 \pm \delta)}^{3}$. The remaining parts of the crystal undergo 
mixed deformation, in which the local finite strain tensor can be 
decomposed into dilatation and pure shear components
\citep{Re41}. The decomposition implies dilatations with volume 
changes by factors of $(1 - \delta)(1 + \delta)(1 \pm \delta)$ 
and the shear strain of about $2\delta$.

\begin{figure}
\noindent
\ifLogReview
    \centering\includegraphics[width=0.75\textwidth]{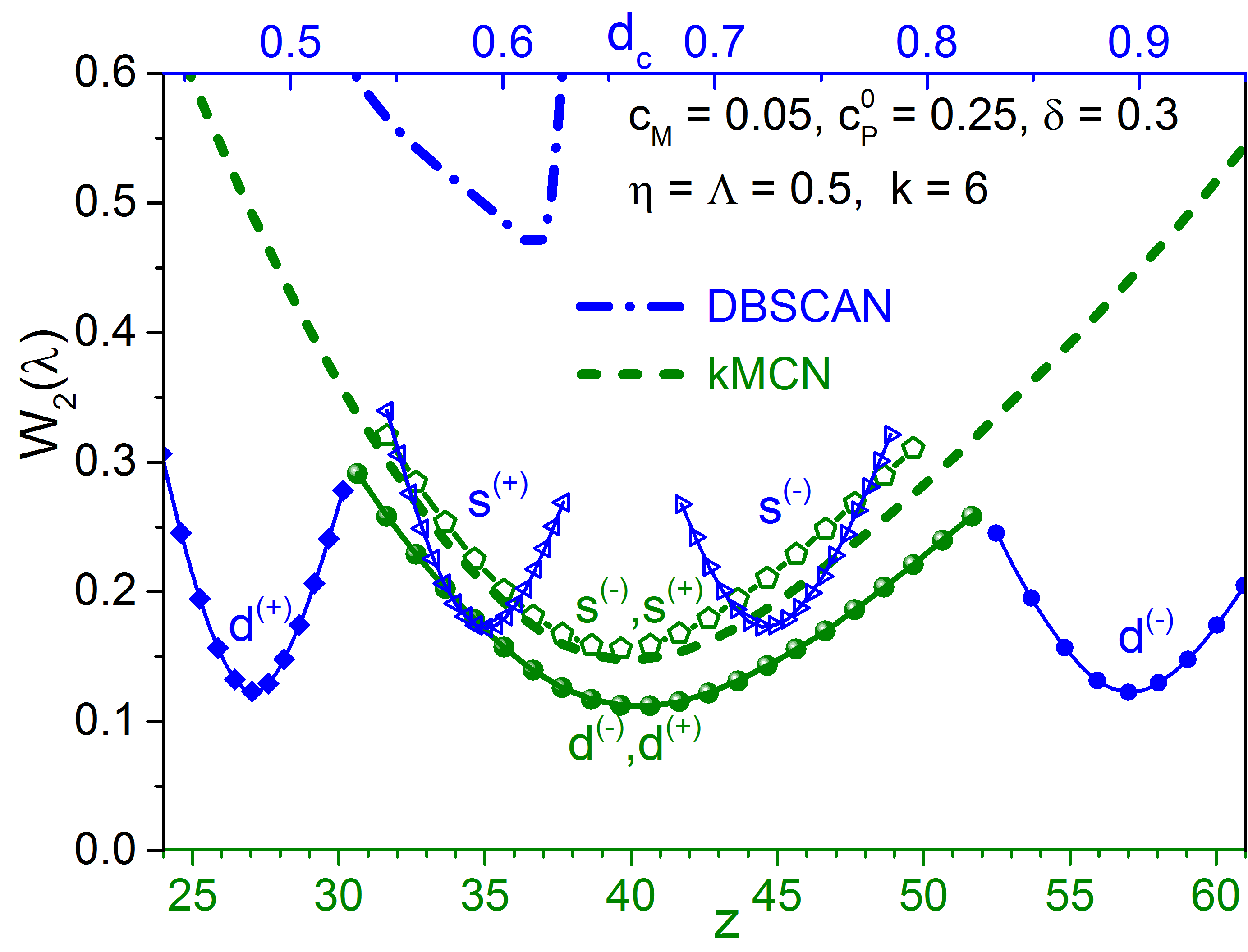}
\else
    \centering\includegraphics[width=0.48\textwidth]{image13.png}
\fi
\caption{
Dependences of clustering quality metrics on control
parameters separately for each of the four types of regions of a
heterogeneously deformed crystal by the field \eqref{EQ_1_} at the strain
$\delta = 0.3$. Filled symbols refer to pure dilatation
($d^{( \pm )}$), open symbols - to regions of mixed deformation
($s^{( \pm )}$). The dashed and dash-dotted lines show the dependencies
for the entire heterogeneously strained crystal.
} \label{fig12}
\end{figure}

Figure~\ref{fig12} shows how the clustering quality metrics $W_{2}$ depend 
on the control parameters $z$ and $d_{c}$ for kMCN and DBSCAN in each
of the specified areas individually. The dilatation component of strain
has no effect on clustering in kMCN (see green circles). The
$W_{2}(z)$ dependence of mixed deformation areas runs slightly higher
(see open green circles), reflecting the contribution of shear
deformations. The resulting green dashed line lies in between. An
important property of these three dependences is that they all reach
their minima at the same value of the control parameter $z$, in this
case around $40$.

DBSCAN yields completely different results (see triangular and rhombic
symbols in Fig.~\ref{fig12}). A clear direct correlation with the densities of
the four areas specified above is evident. The position of the local
minimum $d_{c}^\text{(opt)}$ of the function $W_{2}(d_\text{c})$ is
proportional to $\rho^{- 1/3}$, where $\rho$ is the average atomic
density in the respective area. Due to variations in atomic densities,
the optimal parameter $d_\text{c}^{(opt)}$ in one area is not proper for
another area. Consequently, the DBSCAN dependence $W_{2}(d_\text{c})$
averaged over the entire crystal (dash-dotted line), lies significantly
above the kMCN line. Therefore, it is the variation in dilatation
deformations that leads to a significant degradation of
DBSCAN performance.

Our analysis shows that when using both kMCN and DBSCAN, shear stress
alone has little effect on clustering quality across the entire range of
precipitate compositions and shear strain magnitudes studied. Clustering
quality was assessed using metrics \eqref{EQ_15_}, with kMCN consistently
outperforming DBSCAN.

\subsection*{Superposition of multiple-type precipitates}

So far, we have considered clustering in systems with precipitates of a
given single size $n_\text{P}$ and single composition $c_\text{P}$. Since 
this study does not allow precipitate overlaps, the integral cluster
distributions represent a superposition of the contributions from
individual precipitates. Let us assume that during the simulation we
have constructed a cluster distribution
$\varphi\left( \lambda,n_\text{P},c_\text{P},s \right)$ of a given size
$n_\text{P}$, and composition $c_\text{P}$ for the set of control parameters
$\lambda$. The cluster distribution in the mixed system is then
expressed as a weighted sum of the obtained partial distributions,
\begin{equation} \label{EQ_17_} 
\overline{\varphi}(\lambda,s) = 
\sum_{n_\text{P},c_\text{P}}^{}{\Delta_{n_\text{P}c_\text{P}}\varphi
\left( \lambda\,,n_\text{P},c_\text{P},s \right)},
\end{equation}
where $\Delta_{n_\text{P} c_\text{P}}$ is the fraction of solute atoms in
precipitates of size $n_\text{P}$ and solute concentration $c_\text{P}$.

\begin{figure}
\ifLogReview
    \centering\includegraphics[width=0.75\textwidth]{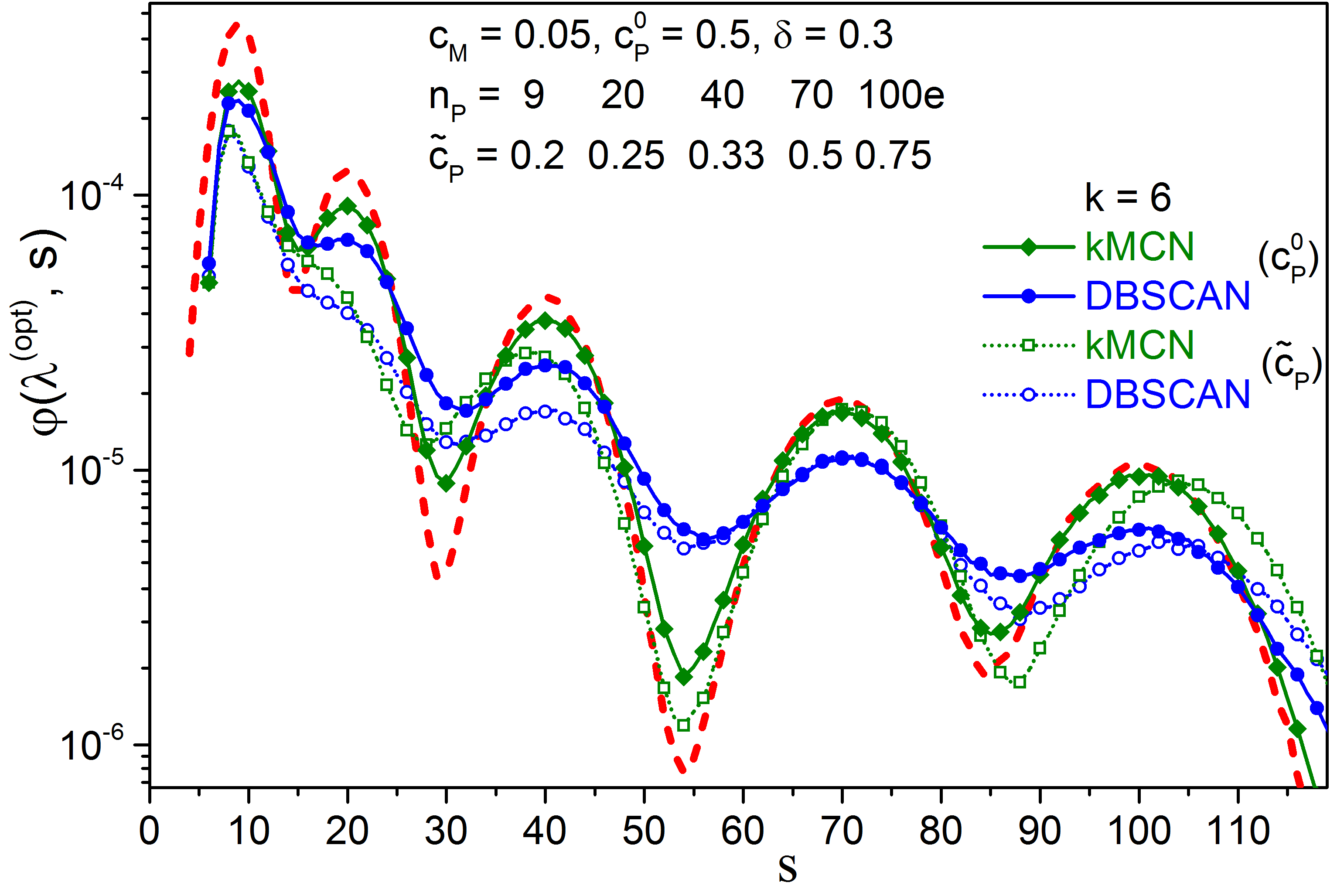}
\else
    \centering\includegraphics[width=0.48\textwidth]{image14.png}
\fi
\caption{
Comparison of the kMCN and DBCSAN methods applied to
heterogeneously deformed structures containing precipitates of different
sizes and compositions. The nominal precipitate sizes are:
$n_\text{P} = 9, 20, 40, 70$ for spherical shape, and $n_\text{P} = 100$
for prolate spheroids. The red dashed line shows the precipitate size
distribution. In a system with precipitates of the same nominal
composition ($c_\text{P}^{0} = 0.5$), the reduced cluster size
distributions obtained at the optimal control parameters
$z^\text{(opt)} = 24$ and $d_{c}^\text{(opt)} = 0.53$~nm are shown by filled
symbols. In a system with precipitates of different compositions
(${\widetilde{c}}_\text{P} = 0.2, 0.25, 0.33,  0.5,  0.75$), the optimal
control parameters were $z^\text{(opt)} = 27$ and
$d_{c}^\text{(opt)} = 0.55$~nm. The corresponding cluster distributions
are shown as open symbols.
} \label{fig13}
\end{figure}

Figure~\ref{fig13} demonstrates optimal cluster distributions 
in a strained alloy containing precipitates of different compositions, 
sizes and shapes. 
The cluster distributions shown were reduced according to \eqref{EQ_16_}.
Optimal parameters were determined by \eqref{EQ_15_}. 
For the same nominal composition of all precipitates 
with $c_\text{P} / c_\text{M} = 10$, 
kMCN reproduces the precipitate size distributions well despite the
substantial strain value $\delta$, see filled diamonds in Fig. 13. The
cluster size distribution yields the correct average value of the
precipitate sizes and the fraction of precipitated solutes. The cluster
distribution is very close to the given precipitate distribution for the
largest sizes ($n_\text{P} > 40$), but the reproduction deteriorates for
smaller sizes. This result is partly explained by the non-optimality of
the chosen control parameter $z = 24$ for small precipitates 
(see Fig. \ref{fig7}).
However, the main reason is the dependence of the reproduction
quality on the precipitate size, even with optimal parameters 
(see Fig.\ref{fig10}).

Heterogeneous strain is always challenging for DBSCAN. In the considered
system it yields a cluster distribution that qualitatively reproduces
the given precipitate distribution, though it underestimates the
fraction of precipitated solutes.

Variations in precipitate composition significantly complicate cluster
analysis, as shown by the open symbols in Fig.~\ref{fig13}. In such a complex
system, a single optimal compromise parameter $z^\text{(opt)}$ is no longer
capable of providing comparable efficiency to that in a system with
single-phase precipitates. In this case, locally density-adaptive
methods such as OPTICS \cite{Re42, Re43}  and hierarchical density-based
cluster analysis \cite{Re44}  can be used. The application of these methods
in the coordination-number metrics will be presented elsewhere.

\section{Conclusions}\label{Section7}

In this study, we extended the concept of the maximum coordination number 
\cite{Re10} to the 
higher-order kMCN and kVCN methods. The first one ranks neighboring atoms by
the Euclidean distance, whereas the second one uses geometric properties of
the Voronoi diagram. Although the second approach may prove more robust
to structural perturbations, both methods yield similar results when
analyzing highly disordered APT structures (see Fig.~\ref{fig6}b).

The correlation function of solute atoms in the ideal solid solution
was exactly described by a negative binomial distribution \eqref{EQ_4_}, 
which constitutes the basis for composition spectrum analysis \eqref{EQ_9_}. 
We estimated the composition and effective fraction of interface layers as
a function of structural disorder and precipitate size, as shown in 
Fig.~\ref{fig2}.

Using the formalized approach based on coordination numbers, we
investigated the percolation properties of APT structures. As shown in
Fig.~\ref{fig3} , the percolation thresholds $p_{c}$ exhibit an approximately
linear dependence of $1 / p_{c}$ on the average coordination
number $z$. Moreover, for high-order methods ($k > 4$) the quantity
$p_{c}z / k$ proves to be almost invariant with respect to the
clustering method parameters $k$ and $z$, as well as the structural
parameters $\Lambda$ and $\eta$. This scaling behavior
motivated the introduction of the self-similar variable \eqref{EQ_13_}, which
provides a relevant description of clustering and enables the direct
transfer of optimal parameters between methods (see Fig.~\ref{fig9}). 
The behavior of clustering curves further clarifies the features of different
clustering methods when applied to complex systems (see Fig.~\ref{fig5}).

Simulations of cluster distributions in a two-phase system showed that
the results depend on the clustering method, as illustrated in 
Figs.~\ref{fig6}--\ref{fig10}.
Methods based on coordination numbers consistently outperformed
density-oriented methods. We identified two main sources of discrepancy
between the calculated cluster distributions and the given precipitate
distributions. The first one appears in the small-cluster regime as
background clustering in the matrix. The second one arises from the
attachment of matrix solutes to clusters within the precipitates (see
Fig.~\ref{fig6}). These precipitate clusters can be effectively separated 
from matrix clusters by subtracting the matrix contribution \eqref{EQ_16_}.

Regarding the order $k$ of the clustering method, the quality of the
cluster analysis improves distinctly as $k$ increases when the
$c_\text{P}/c_\text{M}$ ratio is small. At large $c_\text{P}/c_\text{M}$
ratios, however, the dependence on $k$ becomes negligible 
(see Fig.~\ref{fig10}).

Simulation of local magnification effects demonstrated a clear advantage
of composition-based methods (kMCN) over density-based methods (DBSCAN).
Pure shear was found to affect cluster analysis significantly only at
high strain levels and to depend on precipitate composition. Under these
conditions, kMCN slightly outperformed DBSCAN. The contrast between the
two methods becomes especially pronounced under heterogeneous
dilatation. kMCN is entirely resistant to dilation-type strains, whereas
the performance of DBSCAN degrades rapidly as the amplitude of
heterogeneous dilatation increases (see Figs.~\ref{fig11}~and~\ref{fig13}).


For the cluster analysis of APT data, we recommend replacing methods 
based on the Euclidean metric (e.g., DBSCAN) with their 
coordination-number-based counterparts, such as kMCN, 
which performed better in many cases, never worse, 
and required only a marginal increase in computational cost.


\begin{acknowledgments}
The authors thank N. Wanderka and C. Abromeit for valuable discussions
and critical comments. 
M.L. acknowledges Helmholtz-Zentrum Berlin for financial support and 
for providing access to the computer cluster used for extensive 
simulations.
\end{acknowledgments}

\appendix

\renewcommand{\thefigure}{A.\arabic{figure}}
\setcounter{figure}{0}

\section{The effect of the metric on the configuration of links}
\subsection{Distributions of local neighbor numbers}
We chose the arithmetic mean ($M^\text{A}$) of the
coordination numbers \eqref{EQ_2_} as the metric in this study; 
one could otherwise use the geometric ($M^\text{G}$) 
or harmonic ($M^\text{H}$) mean.
Previously, we used marginal metrics based on backward
($M_{ij}^\text{B} = \min(m_{ij},m_{ji})$ and forward
($M_{ij}^\text{F} = \max(m_{ij},m_{ji})$ symmetrizations \cite{Re10}.
It was demonstrated that the metric $M^\text{F}$ exhibits certain
advantages in regions of abrupt change in atomic density, such as
external surface and internal extended defects. Metric \eqref{EQ_2_} 
preserves this property, as the inequalities
$M^\text{F} \geq M^\text{A} \geq M^\text{G} \geq M^\text{H} \geq M^\text{B}$ 
hold, and $M^\text{A}$ is closest to $M^\text{F}$. 
The simplicity of computing metric
\eqref{EQ_2_}, which uses only integer arithmetic, is also important.
\begin{figure}
\ifLogReview
    \centering\includegraphics[width=0.75\textwidth]{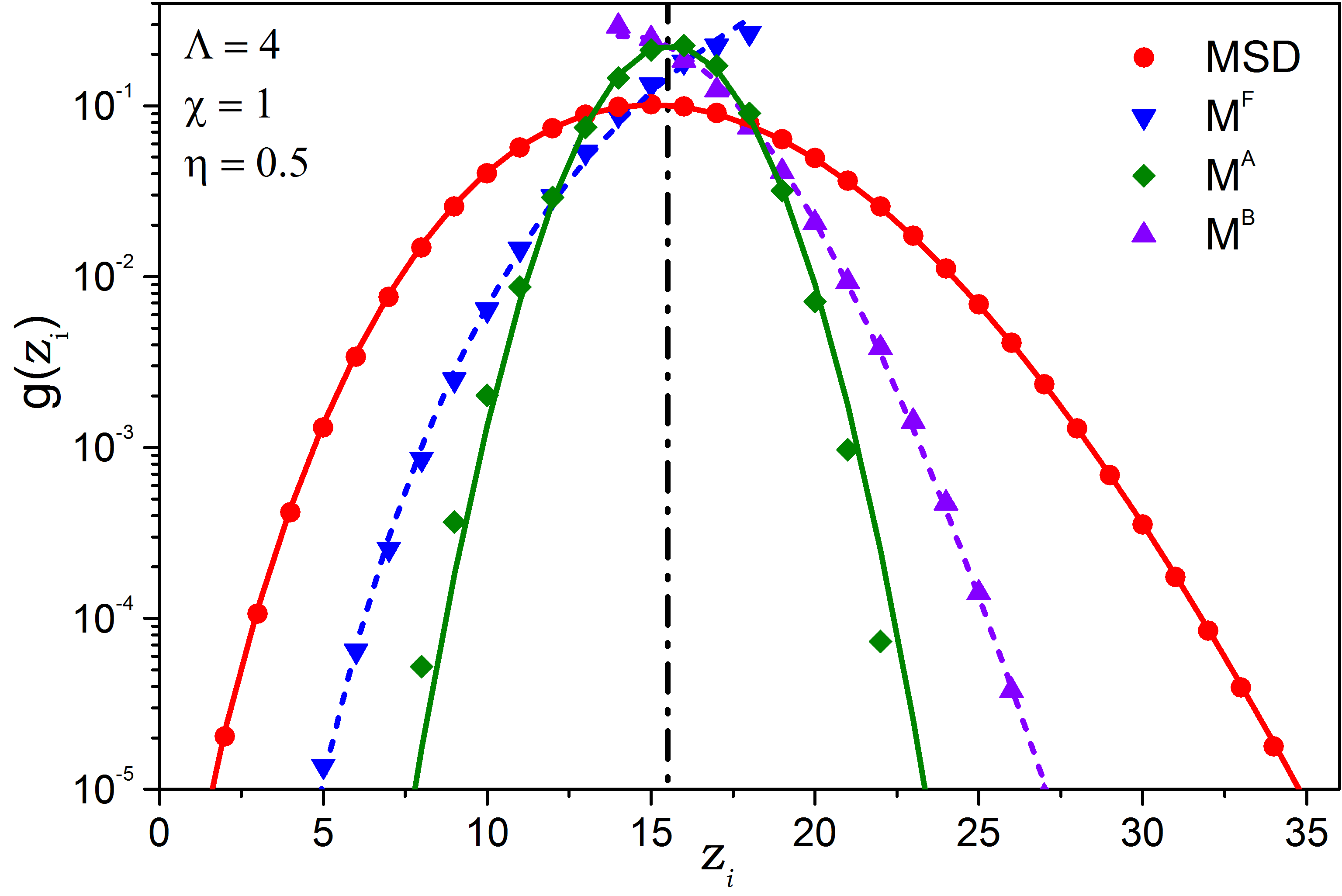}
\else
    \centering\includegraphics[width=0.48\textwidth]{image15.png}
\fi
\caption{
The nearest-neighbor number distributions in MSD
and variants of the MCN metric, using a gas-like random structure. The
parameter $z_{0}$ was set to 18, 16 and 14 for the
$M^\text{F}, M^\text{A}$ and $M^\text{B}$
methods, respectively. The dash-dot line shows the location of the mean
$z \approx 15.5$ for all distributions. The solid lines: fittings of MSD
by the Poisson and $M^\text{A}$ by the Gaussian distributions
respectively. The dashed lines: fittings of $M^\text{F}$ and
$M^\text{B}$ by two-parameter Gamma distributions.
} \label{figA1}
\end{figure}

Figure~\ref{figA1} compares the distributions of local neighbor numbers
$z_{i}$ for different metrics in a gas-like system.
The number of neighbors within the Euclidean separation
$d_{c}$ follows the Poisson distribution with the variance equal to
$z(d_{c})$, as defined by \eqref{EQ_3_}. 
The marginal metrics $M^\text{F}$ and
$M^\text{B}$ produce highly skewed distributions. The MCN method using
metric \eqref{EQ_2_} yields an almost symmetric distribution with the least
variance, which can be roughly estimated by the relation
$\sigma^{2} \approx 1 + 0.12z$. Apparently, the narrowness of this
distribution provides some advantages for the MCN method in cluster
analysis.

To further reduce the variance, we can redefine the coordination numbers
$m_{ji}$ based on the coordination distances \eqref{EQ_2_}, rather than the
Euclidean distances used initially. The first iteration of the
redefinition reduces the variance by a factor of 2-3. Each subsequent
iteration narrows the distribution even more. This procedure provides an
easy alternative to constructing a topologically disordered lattice with
a constant coordination number \cite{Re45}.

\begin{figure}
\ifLogReview
    \centering\includegraphics[width=0.75\textwidth]{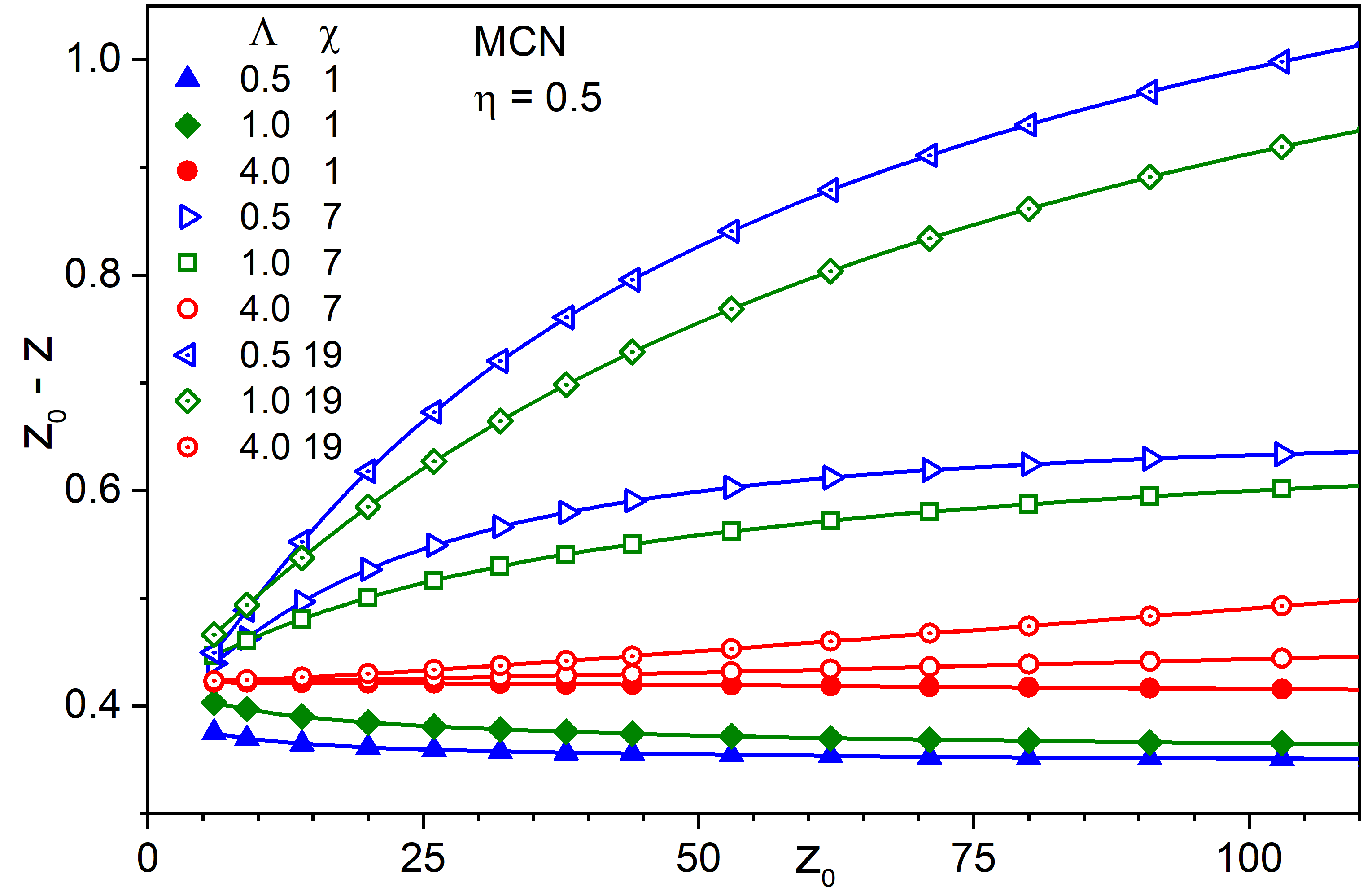}
\else
    \centering\includegraphics[width=0.48\textwidth]{image16.png}
\fi
\caption{
Dependence of the average number of neighbors $z$
on the MCN parameter $z_{0}$ for the metrics \eqref{EQ_2_} at different values
of spatial resolution $\Lambda$ and correlated loss $\chi$.
} \label{figA2}
\end{figure}

The dependences of the mean $z$ of the neighbor distribution on the
parameter $z_{0}$ and the properties of structural disorder obtained
within MCN and metric \eqref{EQ_2_} are shown in Fig.\ref{figA2}. 
It appears that $z$
is always slightly less than $\, z_{0}$. For the uncorrelated atom
loss ($\chi = 1$), the difference $z_{0} - z$ does not exceed about
0.4 both in the APT-like structure ($\Lambda = 0.5 - 1$) and for the
gas-type disorder ($\Lambda > 4$). For the correlated atom loss
($\chi > 1$), it increases but only slightly.

\subsection{Maps of links}
\begin{figure*}
\ifLogReview
\centering (a)\hspace{14em} (b)\hspace{14em} (c)\\
\centering\includegraphics[width=\textwidth]{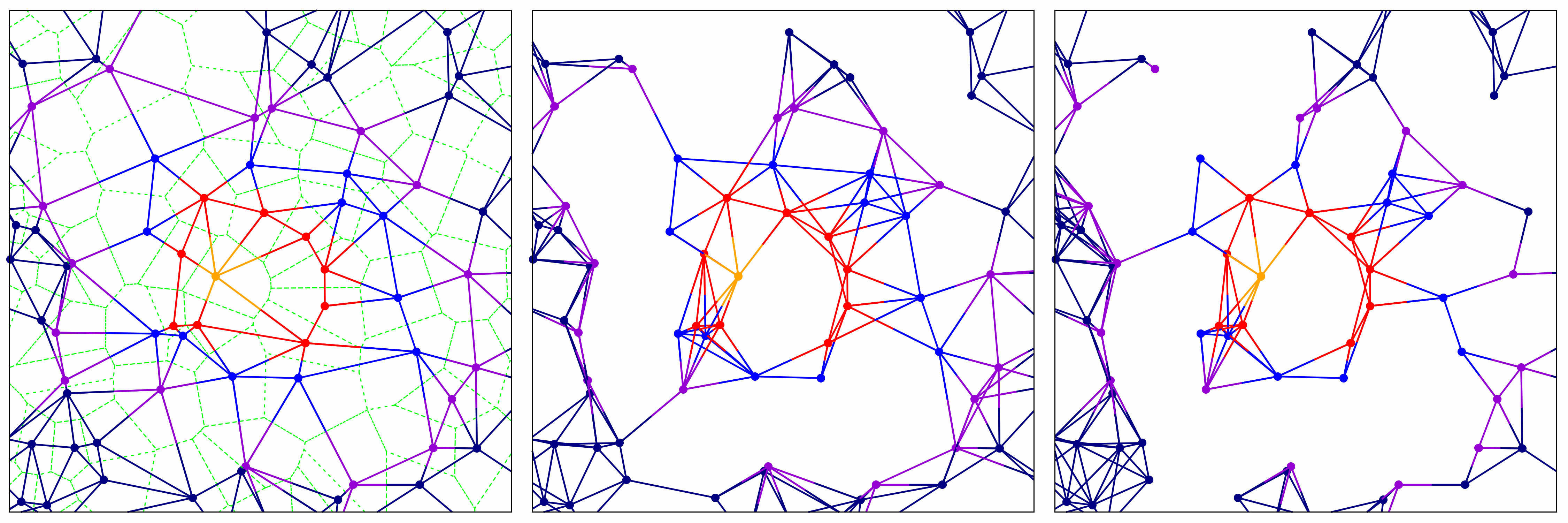}
\else
\centering (a)\hspace{15em} (b)\hspace{15em} (c)\\
\centering\includegraphics[width=0.9\textwidth]{image17.png}
\fi
\caption{
The maps of links (solid lines) obtained in various metrics
for the atomic arrangement shown in Fig.~\ref{fig1}. The line colors 
represent the shells of the central atom. 
(a) VCN metric: The dashed lines depict
the Voronoi diagram. The control parameter $z_{0}$ was set to $6$,
resulting in the average number of links $z\,$ of about $5.4$. 
(b) MCN metric: The same control parameter $z_{0}$, resulting in
$z \approx 5.6$. 
(c) MSD metric: The parameter $d_{c}$
corresponds to $z$ = 6 according to Eq.~\eqref{EQ_3_}.
} \label{figA3}
\end{figure*}

To gain a visual understanding of metric properties, it is helpful to
compare the maps of links that form in a highly disordered system. To
this end, we consider the same fragment of 2D atomic arrangement shown
in Fig.~\ref{fig1} as a reference system. Figure~\ref{figA3} demonstrates
the maps obtained by three methods with the control parameter $z_{0} = 6$,
equal to the average number of edges per Voronoi cell for a homogeneous
Poisson point process in 2D.

The VCN graph (Fig.~\ref{figA3}a) appears to be the most balanced. 
Almost all atoms have 5-6 nearest neighbors. 
In contrast, the MSD graph (Fig.~\ref{figA3}c) is the most unbalanced 
one. A distinct crowding of links is clearly visible on
the left side, and sparseness on the right side. 
The MCN graph (Fig.~\ref{figA3}b)
appears intermediate with an acceptable balance. It should be noted that
the comparison of methods was chosen at the most favorable control
parameter value for VCN: $z_{0} = 6$. As $z_{0}$ increases, the
properties of the VCN and MCN graphs converge. In 3D, we observed some
advantages of the VCN in the range of about $z_{0} \approx (12 - 20)$.

\renewcommand{\thefigure}{B.\arabic{figure}}
\setcounter{figure}{0}

\section{Simulation of percolation thresholds in high-order methods}
To estimate the percolation threshold, the simulation proceeds in three
stages: 
(a) a random site structure is generated in a box of
linear size $\mathcal{L}$ with disorder parameters $\eta$,
$\Lambda$, and $\chi$; and neighbor lists are constructed for each
site using $z_{0}$ to identify core points and $z_{L}$ to define
links between them; 
(b) a modified Newman--Ziff algorithm is then used
to calculate the statistical quantity $P_{n}$ in the microcanonical
ensemble, with an additional check after each atom insertion to detect
newly formed core points, while clustering is performed only on core
points; and (c) after many repetitions of the previous steps, a
convolution-type transformation is applied to obtain $P(c)$ in the
canonical ensemble \cite{Re37}.

This algorithm yields the size of the largest cluster directly, so we
use this quantity, normalized by the total number of sites, as $P(c)$.
It is well established that the scaling behavior of the largest cluster
size near the percolation threshold is given by \cite{Re46}:
\begin{equation} \label{EQ_B1_} 
P_\text{max} (c, \mathcal{L} ) \sim 
\mathcal{L}^{-\beta / \nu}\mathcal{F}\left( (c - p_c)\mathcal{L}^{1/\nu} \right)
\end{equation}
where $\mathcal{F}$ is a universal scaling function for systems with
identical configurations and boundary conditions. In an ideal lattice,
any box size $\mathcal{L}$ satisfies these requirements. In a random
structure, however, $\mathcal{L}$ must be large enough to ensure
statistical equivalence among systems of different sizes. For the same
reason, a random structure is generated multiple times for sampling.
For a typical box size of $\mathcal{L} = 100$, we use at least
$N_{\text{conf}} = 250$ structure realizations (Stage 1) and more
than $N_\text{iter} = 10^{5}$ iterations (Stage 2). The ratio of critical
exponents $\beta / \nu$ depends only on the dimension of space
$d$ \cite{Re35}. It is equal to $d - d_{f}$, where $d_{f}$ is the
fractal dimension of the percolating cluster, which is approximately
$2.523$ for $d = 3$ \cite{Re47}.

\begin{figure}
\ifLogReview
    \centering\includegraphics[width=0.75\textwidth]{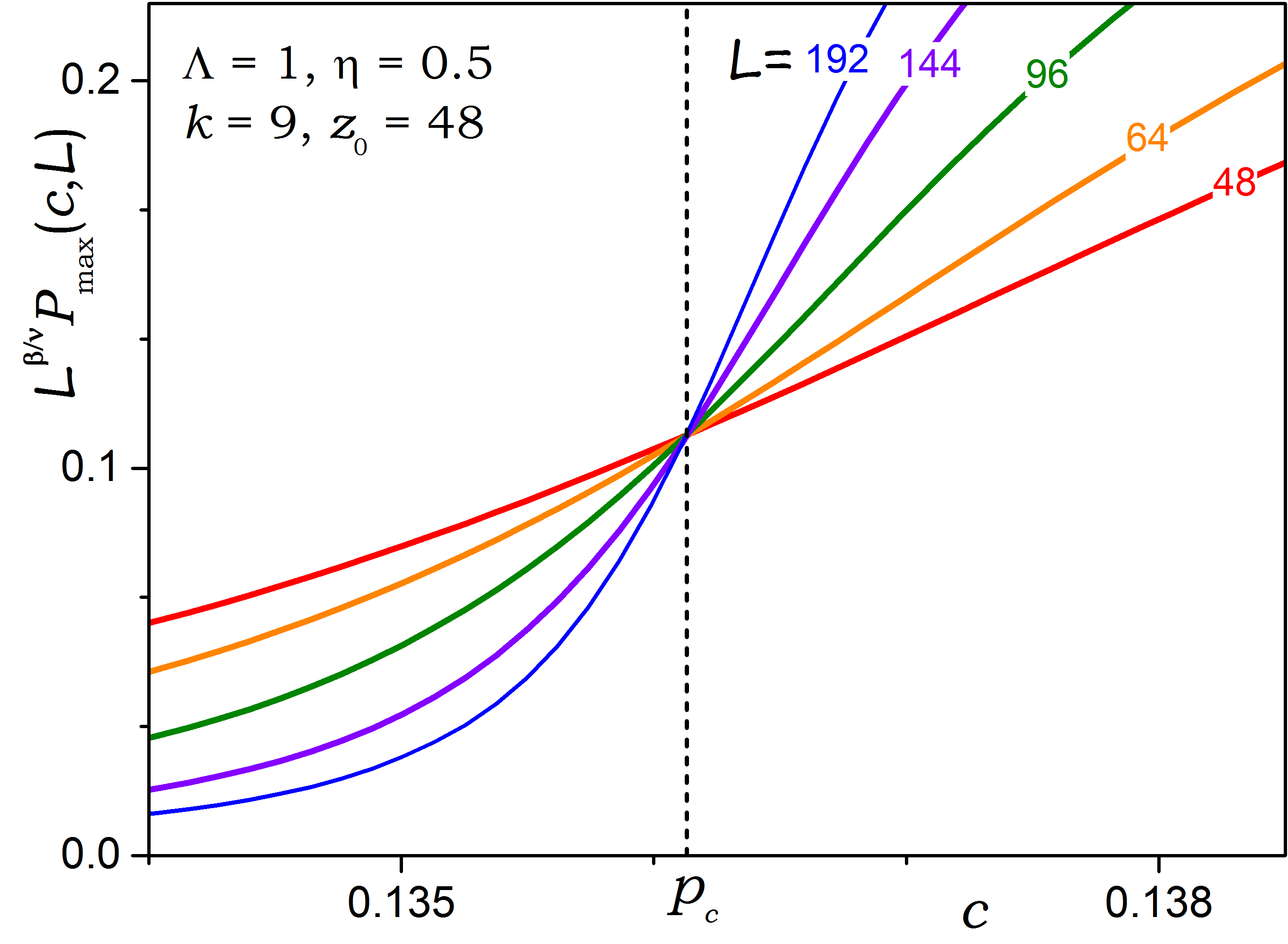}
\else
    \centering\includegraphics[width=0.48\textwidth]{image18.png}
\fi
\caption{
Dependence of
$\mathcal{L}^{\beta / \nu} P_{\text{max}} (c, \mathcal{L})$ 
on the concentration $c$ and the box size $\mathcal{L}$ for 
high-order ($k = 9$) percolation. The scaling parameter used is
$\beta / \nu = 0.477$ \cite{Re48}. 
The calculated coordination-number mean is $z = 47.627$ at $z_{0} = 48$.
} \label{figB1}
\end{figure}

It follows from ~\eqref{EQ_B1_} that the curves
$\mathcal{L}^{-\beta/\nu}P_{\text{max}}\left( c, \mathcal{L} \right)$
intersect near $c = p_{c}$ for any valid values of $\mathcal{L}$.
Figure~\ref{figB1} illustrates an application of this approach. Under the
conditions indicated in the plot, the obtained percolation threshold
$p_{c}(k,z)$ is $0.13613(2)$. 
For comparison, formulas ~\eqref{EQ_10_} and ~\eqref{EQ_11_},
using the parameters listed in Table 1, yield values of $0.13622$ and
$0.13612$, respectively.

\section{Nomenclature}
\textbf{Main parameters of the simulated alloy}

$n_\text{P}$ -- precipitate size (numbers of solutes 
after random atomic removal)

$L$ -- lattice constant

$\Lambda$ -- atomic coordinate resolution in lattice constant units

$\eta$ -- detection efficiency

$\chi$ -- correlated loss parameter

$\delta$ -- the strain parameter

$c_\text{M}$ -- matrix composition 
(concentration of solutes in the matrix phase)

$c_\text{P}$ -- precipitate composition 
(concentration of solutes in the precipitate phase)

$\Delta_\text{P}$ -- fraction of solutes in the precipitates

\textbf{Clustering methods}

MSD -- Maximum Separation Distance

MCN -- Maximum Coordination Number

VCN -- MCN with coordination numbers based on Voronoi diagram

DBSCAN -- Density-Based Spatial Clustering of Applications with Noise
($k$-order MSD)

kMCN -- $k$-order MCN

kVCN -- $k$-order VCN

\textbf{Main parameters of clustering methods}

$k$ -- order of clustering method

$d_{c}$ -- control parameter of MSD and DBSCAN

$m_{ji}$ -- coordination number defined as the sequence number of an
atom $j$ in the sorted neighbor list of an atom $i$

$M_{ij} = \left( m_{ij} + m_{ji} \right) / 2$ -- coordination
distance between atoms $i$ and $j$

$z_{0}$ -- control parameter of MCN and kMCN

$z$ -- average number of neighboring atoms at the fixed $z_{0}$.

$z^\text{(opt)}$ -- optimal control parameter

$s_{m}$ -- minimal cluster size 
(minimum accountable number of solute atoms in the cluster)

$\lambda$ -- set of main parameters i.e $k,z_{0},s_{m}$ 
for the kMCN method

\textbf{Other nomenclature}

$g_\text{rdf}(r)$ -- the radial distribution function

$\rho_{0} = N /V$ -- the average atomic density

$N$ -- the total number of atoms

$V$ -- the volume of the system

$N_\text{B}$ -- the number of solute atoms of the system

$B\left( m \middle| k,c \right)$ -- negative binomial distribution,
Eq. \eqref{EQ_4_}

$p_{c}$ -- the percolation threshold

$A(k)$ and $a(k)$ -- parameters of percolation threshold linear
approximation \eqref{EQ_11_}

$\zeta(k,z,c)$ -- self-similar variable \eqref{EQ_13_}

$\varphi(s,\zeta)$ -- the size ($s$) distribution of 
solute clusters or precipitates

$F\left( s_{m},\zeta \right)$ -- clustering curve or the fraction of
clustered solutes Eq.~\eqref{EQ_14_}

$W_{p}(\lambda)$ -- statistical distance or measure 
of optimality Eq.~\eqref{EQ_15_}


\bibliography{APT2P}

\end{document}